\documentclass[aip,amsmath,amssymb,reprint,groupedaddress]{revtex4-2}
\usepackage{graphicx}% Include figure files
\usepackage{dcolumn}% Align \mbox{TABLE} columns on decimal point
\usepackage{bm}% bold math
\usepackage[utf8]{inputenc}
\usepackage[T1]{fontenc}
\usepackage{mathptmx}

\usepackage{siunitx}
\DeclareSIUnit\au{a.u.}
\DeclareSIUnit\atomicunits{a.u.}
\DeclareSIUnit\radpi{rad}

\usepackage{xfrac}

\usepackage[version=4]{mhchem}

\usepackage{physics}

\usepackage{multirow}

\usepackage{subfigure}

\usepackage{xr}

\usepackage{dcolumn}% Align \mbox{TABLE} columns on decimal point
\usepackage{bm}% bold math

\begin{document}

\preprint{AIP/123-QED}

\title[]{Coupled nuclear and electron dynamics in the vicinity of a conical intersection}
% Force line breaks with \\

\author{Thomas Schnappinger}
\affiliation{Department of Chemistry, LMU Munich, Germany, D-81377 Munich, Germany}
\author{Regina de Vivie-Riedle}
\affiliation{Department of Chemistry, LMU Munich, Germany, D-81377 Munich, Germany}

\date{\today}% It is always \today, today,

\begin{abstract}
Ultrafast optical techniques allow to study ultrafast molecular dynamics involving both nuclear and electronic motion. To support interpretation, theoretical approaches are needed that can describe both the nuclear and electron dynamics. Hence, we revisit and expand our ansatz for the coupled description of the nuclear and electron dynamics in molecular systems (NEMol). In this purely quantum mechanical ansatz the quantum-dynamical description of the nuclear motion is combined with the calculation of the electron dynamics in the eigenfunction basis. The NEMol ansatz is applied to simulate the coupled dynamics of the molecule \ce{NO2} in the vicinity of a conical intersection (CoIn) with a special focus on the coherent electron dynamics induced by the non-adiabatic coupling. Furthermore, we aim to control the dynamics of the system when passing the CoIn. The control scheme relies on the carrier envelope phase (CEP) of a few-cycle IR pulse. The laser pulse influences both the movement of the nuclei and the electrons during the population transfer through the CoIn. 
\end{abstract}

\maketitle

%\begin{quotation}
%The ``lead paragraph'' is encapsulated with the \LaTeX\ 
%\verb+quotation+ environment and is formatted as a single paragraph before the first section heading. 
%(The \verb+quotation+ environment reverts to its usual meaning after the first sectioning command.) 
%Note that numbered references are allowed in the lead paragraph.
%
%The lead paragraph will only be found in an article being prepared for the journal \textit{Chaos}.
%\end{quotation}

\section{\label{sec:Introduction}Introduction}
The continuous development of attosecond laser pulses enables spectroscopic techniques which allow the time resolved investigations of ultrafast photo-initiated processes in atoms, molecules and solids. Nowadays it is possible to study electronic correlation and ultrafast molecular dynamics through pump-probe experiments\cite{Hentschel:2001,Goulielmakis:2007,Sansone:2010,Kraus:2013,AndreaRozzi:2013,Romero:2014,Calegari:2014}. Within these experiments attosecond, broad-band pulses are used to generate electron wavepackets in highly excited states of molecules, leading to the discovery of effects such as electron localization in diatomic molecules\cite{vondenHoff:2009,Sansone:2010} and, later, of purely electronic charge migration in biological relevant molecules\cite{AndreaRozzi:2013,Romero:2014,Calegari:2014}. 
To explain and interpret the observations of these experiments, theoretical approaches are needed that can describe the dynamics of electrons in molecules. Most approaches use time-dependent analogs of well-established quantum-chemical methods like time-dependent Hartree-Fock theory (\mbox{TD-HF})\cite{Kulander:1987} or time-dependent density-functional theory (\mbox{TD-DFT})\cite{Runge:1984}. Furthermore, time-dependent post-Hartree-Fock methods like time-dependent configuration-interaction (\mbox{TD-CI})\cite{Klamroth:2003,Rohringer:2006}, time-dependent coupled-cluster (\mbox{TD-CC})\cite{Skeidsvoll:2020,Vila:2020} and multi-configuration time-dependent Hartree-Fock\cite{Zanghellini:2004} are available for the correlated description of electron dynamics in molecular systems. In other theoretical approaches the electronic wavefunction is propagated directly in time, with the help of Green's function\cite{Kuleff:2005} or in the basis of molecular orbitals\cite{Remacle:2006}. All these theories focus on the evolution of the electronic subsystem, driven by electronic correlation\cite{Kuleff:2007,Golubev:2015} and predict long-lived coherences. The neglect of the nuclear motion is justified by the assumption that the dynamics of the electrons is much faster than the one of the heavier nuclei. This results in charge migration, an oscillatory motion of electron density with frequencies defined by the energy gaps among the states populated with the initial laser pulse. If the states of the superposition are close together, the electron dynamics becomes slow and therefor the nuclear motion can no longer be neglected. But as shown in numerous theoretical works\cite{vondenHoff:2009,vondenHoff:2012,vondenHoff:2012:2,Vacher:2015,Jenkins:2016,Arnold:2017,Dongming:2019}, nuclear motion in general causes decoherence in molecular systems and should not be neglected in no cases. This decoherence causes the electronic wavepackets to  exist only for short time scales\cite{Arnold:2017}. For small systems like \ce{H2+} or \ce{D2+} a full quantum treatment of the coupled electron and nuclear dynamics is possible\cite{Bandrauk:2004}. Beyond these three particle problems there are computationally very demanding methods available based on a multi-configurational ansatz\cite{Nest:2009} or on the coupled description of nuclear and electronic ﬂux\cite{Vincent:2016,Matsuzaki:2019}. Further techniques are based on the coupled propagation of the nuclear and electronic wavefunction on a single time-dependent potential energy surface\cite{Cederbaum:2008,Abedi:2010,Abedi:2012,Chiang:2014}. But for larger molecular systems the main techniques used are mixed quantum classical representations\cite{Ojanperae:2012,Alonso:2012,Takatsuka:2017,Takatsuka:2018}. For example, the electron dynamics is described using TD-DFT and the nuclear motion is considered using an Ehrenfest approach\cite{Ojanperae:2012,Alonso:2012}. But these methods do not reflect the quantum nature of the nuclei which, however, becomes important for ultrashort pulse excitation and non-adiabatic transitions.

In this paper we want to revisit and expand an ansatz for the coupled description of the nuclear and electron dynamics in molecular systems\cite{Geppert:2008,vondenHoff:2009,Znakovskaya:2009} (shortened NEMol) developed in our group. It is based on electronic structure calculations as well as nuclear quantum dynamics. In its initial formulation the electronic wavefunctions are represented as Slater determinants and propagated in the eigenstate basis. The coupling of the nuclear motion to the electron motion is incorporated explicitly through the nuclear wavepacket motion as well as through a coherence term with contributions from the nuclear and electronic wavefunctions. Compared to the similar approaches\cite{Cederbaum:2008,Abedi:2010,Abedi:2012,Chiang:2014}, the feedback of the electron motion to the nuclear dynamics is less directly introduced by simulating the nuclear dynamics on coupled potential energy surfaces (PES).
The central equation of the original NEMol ansatz\cite{Geppert:2008,vondenHoff:2009,Znakovskaya:2009} relates the dynamics of the coupled one-electron density to the temporal evolution of the expected value of the nuclear positions. In the first part of this work we want to generalize the NEMol ansatz by extending beyond this single geometry approximation. Therefore, we introduce the \mbox{NEMol-grid} in order to represent the electron dynamics at multiple points on the grid used for the nuclear wavepacket propagation. In the limit the NEMol-grid is equal to the grid representing the nuclear wavepacket, but in practice we choose a coarser one. By means of a simple approximation it is possible to obtain a condensed representation of time-dependent electron density in the one-electron-two-orbital (1e-2o) picture.

In the second part we want to explore the potential of our NEMol ansatz. For this purpose, we consider a situation that can generate coherent electron dynamics in excited states of molecules even without a laser pulse present. Such a scenario occurs in the vicinity of a conical intersection (CoIn)\cite{Domcke:2004,Abe:2006,Egorova:2008,Kowalewski:2015,Keefer:2020}. For this ubiquitous but nevertheless extraordinary points in a molecular system the adiabatic separation between nuclear and electronic motion breaks down\cite{Yarkony:1996,Baer:2002,Domcke:2004} and the electronic states involved become degenerate. Beside the creation of funnels for radiationless electronic transitions a coherent electron wavepacket is created whose dynamics approaches the time scale of the nuclear dynamics. All these properties of CoIn's are determined by the shape and size of the non-adiabatic coupling elements (NAC's) and the topography of the vicinity. As a realistic molecular system which provides such a situation we have chosen the \ce{NO2} molecule. After excitation into the first excited state a CoIn enables an ultrafast non-adiabatic transition back to the ground state within less than \SI{100}{\femto\second}. This fast relaxation as well as the photophysics of \ce{NO2} in general have been widely explored both theoretically\cite{Haller:1985,Mahapatra:1999,Santoro:1999,Mahapatra:2000,Santoro:2000,Kurkal:2003,Sanrey:2006,Arasaki:2007,Schinke:2008,Arasaki:2010,Richter:2015,Woerner:2018,Richter:2019} and experimentally\cite{Eppink:2004,Form:2006,Vredenborg:2008,Arasaki:2010,Wilkinson:2010,Arasaki:2011,Woerner:2011,Ruf:2012,Woerner:2012}. Beside the free relaxation of \ce{NO2} we also studied the influence on the coupled electron dynamics when applying a few-cycle IR laser pulse in the vicinity of the CoIn. The variation of the carrier envelope phase $\phi$ (CEP) of such a few-cycle pulse offers the possibility to steer electrons and nuclei\cite{Weitzel:2007,Roudnev:2007,Kling:2008,Arasaki:2010,Arasaki:2011,Znakovskaya:2011,Znakovskaya:2012,Kling:2013,Alnaser:2014,Richter:2015,Arnold:2018,Richter:2019,Schueppel:2020}. Similar to previous studies\cite{Arasaki:2010,Arasaki:2011,Richter:2015,Richter:2019} we apply this CEP-control-scheme to \ce{NO2} and evaluate the CEP-dependence of the resulting coupled nuclear and electron dynamics.

\section{\label{sec:NEMol} Coupled Nuclear and electron dynamics (NEMol)}
In the original NEMol ansatz\cite{Geppert:2008,vondenHoff:2009,Znakovskaya:2009} the coupled one-electron density $\rho(r,t;\langle R \rangle(t))$ is defined according to equation~\ref{eq:coupledDens}. For convenience the detailed derivation of this equation can be found in the appendix adapted to the current notation.
\begin{align}
\begin{split}
\label{eq:coupledDens}
\rho(r,t;\langle R \rangle(t)) & =  \sum_j  A_{jj}(t)\rho_{jj}(r;\langle R \rangle(t))  \\
  & + \sum_{k \neq j } 2 Re \big\{  A_{jk}(t)\rho_{jk}(r;\langle R \rangle(t))e^{-i \xi_{jk}(t)} \big\},
\end{split} \\
\text{with } \xi_{jk}(t) & =  \Delta E_{jk}(\langle R \rangle(t)) \Delta t + \xi_{jk}(t - \Delta t ).
\end{align}
The first summation consists of the state specific electronic density $\rho_{jj}(r,t;\langle R \rangle(t))$ weighted with the corresponding time-dependent population $A_{jj}(t)$. The second summation defines the coherent contribution to the coupled electron density and consists of the time-dependent overlap $A_{jk}(t)$, the one-electron transition density $\rho_{jk}(r,t;\langle R \rangle(t))$ and its pure electronic phase defined by the energy difference $\Delta E_{jk}$ between the electronic states involved. All quantities related to the electronic wavefunction are calculated for one nuclear geometry per time step which is defined by the time-dependent expected value of the position~$\langle R \rangle(t)$ (for definition see the appendix). As long as we are focusing on situations with quite localized wavepackets and/or one-dimensional systems\cite{Geppert:2008,vondenHoff:2009,Znakovskaya:2009} this approximation works quite well. But in order to treat higher dimensional systems and more complex processes we want to generalize the NEMol ansatz in this work. To extend the ansatz the integration over the full nuclear coordinate space is split up in segments to improve the resolution of the spatial dependence of  the electronic phase term. For this purpose a second grid, the NEMol-grid, is introduced. The resulting modified NEMol ansatz is described in the following section using exemplary a system with two nuclear coordinates $c_1$ and $c_2$. The complete two-dimensional coordinate space is split up into $M \times L$ segments defined by their boundaries $m_{min}$, $m_{max}$ and $l_{min}$, $l_{max}$. For each of these segments $ml$ the population terms $\alpha_{jj}^{ml}(t)$ and the overlap terms $\alpha_{jk}^{ml}(t)$ are calculated.
\begin{equation}
\label{eq:gridpopover}
\alpha_{jk}^{ml}(t)  =  \int_{m_{min}}^{m_{max}} \int_{l_{min}}^{l_{max}} \chi_{j}^{*}(R,t)\chi_{k}(R,t) dc_{1}dc_{2}.
\end{equation}
The sum of these segment terms results in the corresponding total population and overlap.
\begin{equation}
\label{eq:gridover_sum}
\sum_{m = 1}^M \sum_{l =1}^L \alpha_{jk}^{ml}(t) = \bra{\chi_{j}(R,t)}\ket{\chi_{k}(R,t)}_{R} = A_{jk}(t).
\end{equation}
At the center $R_{ml}$ of each segment the state specific electronic densities, the one-electron transition densities and the eigenenergies are determined and with these values the coupled one-electron density for each segment $\rho_{ml}(r,t; R_{ml})$ is calculated.
\begin{align}
\begin{split}
\label{eq:grid_dens}
\rho_{ml}(r,t; R_{ml}) &=  \sum_j  \alpha_{jj}^{ml}(t) \rho_{jj}(r;R_{ml})\\
  & + \sum_{k \neq j } 2 Re \big\{  \alpha_{jk}^{ml}(t) \rho_{jk}(r;R_{ml})e^{-i \xi_{jk}^{ml}(t)} \big\}, 
  \end{split} \\
\text{with }  \xi_{jk}^{ml}(t) & =  \Delta E_{jk}(R_{ml}) \Delta t + \xi_{jk}^{ml}(t - \Delta t ).
\end{align}
It should be noted that for each segment the $\Delta E_{jk}$ values and the electron densities are no longer dependent on~$\langle R \rangle(t)$. In contrast to the original NEMol ansatz, now many $\Delta E$ values are simultaneously contributing to the overall electron dynamics. They are addressed, whenever the nuclear wavepacket is located there. To obtain the total coupled electron density the individual contributions of each segment are summed up. 
\begin{equation}
\label{eq:gridDens}
\rho(r,t;R)  =  \sum_{m = 1}^M \sum_{l =1}^L  \rho_{ml}(r,t;R_{ml}).
\end{equation}
This total coupled electron density $\rho(r,t;R)$ describes the electron dynamics coupled to multiple grid points on which the nuclear wavepacket is represented.

A second aspect that we would like to introduce is a further simplification. For clarity reasons it is here  formulated in terms of the original NEMol ansatz. We now consider a system of two electronic states described by their electronic wavefunctions $\varphi_{1}$ and $\varphi_{2}$. In the simplest case the wavefunctions of both states are described by two Slater determinants which only differ in the occupation of one spin orbital $\theta$. Now the coupled total electron density can be simplified by expressing the densities and transition densities using the spin orbitals.
\begin{widetext}
\begin{equation}
\rho(r,t;\langle R \rangle(t))  =  \sum_{j=1}^{N-1} |\theta_{j}(r;\langle R \rangle(t))|^{2}   + \sum_{k=1}^2 A_{kk}(t) |\theta_{k}(r;\langle R \rangle(t))|^2 +  2 Re \big\{ A_{12}(t) \theta_{1}(r;\langle R \rangle(t))\theta_{2}(r;\langle R \rangle(t)) e^{-i\xi_{12}(t)} \big\}.
\label{eq:simpelcoupledDens}
\end{equation}
\end{widetext}

The summation at the beginning includes the densities of all equally occupied orbitals and is followed by the densities of the remaining two orbitals $\theta_{1}$ and $\theta_{2}$ weighted with the populations $A_{11}(t)$ and $A_{22}(t)$ The coherent part contains the product of the orbitals $\theta_{1}$ and $\theta_{2}$. Within this simplification it is now possible to neglect the contributions of the equally occupied orbitals in order to study the coupled electron dynamics in an one-electron-two-orbital (1e-2o) picture. Under the above mentioned approximation this 1e-2o picture is a possibility to examine the coherent part of the electron dynamics in a very condensed way. This simplification can also be made in combination with the NEMol-gird.

\section{\label{sec:no2_dyn} \ce{NO2} coupled dynamics}
\begin{figure}[ht!]
    \centering
    \subfigure[][]{
        \includegraphics[width=0.45\textwidth]{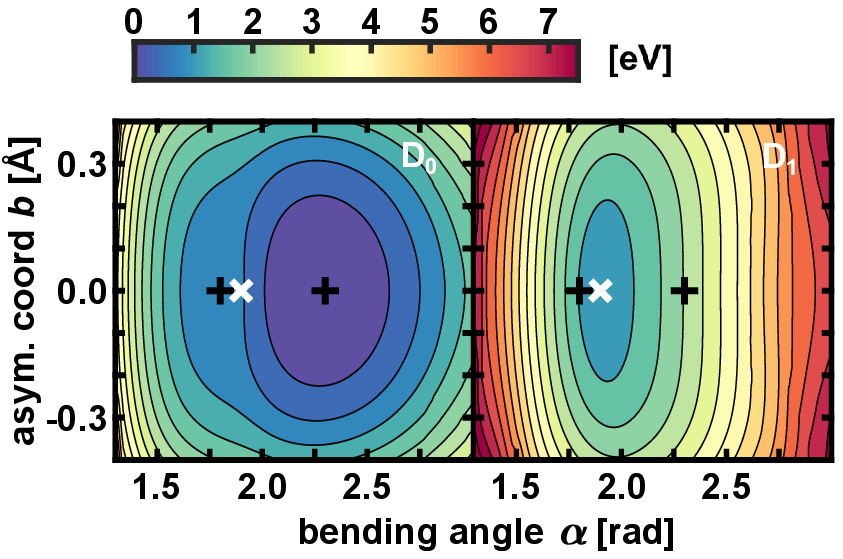}
        \label{fig:Pots}
    }
    \subfigure[][]{
        \includegraphics[width=0.45\textwidth]{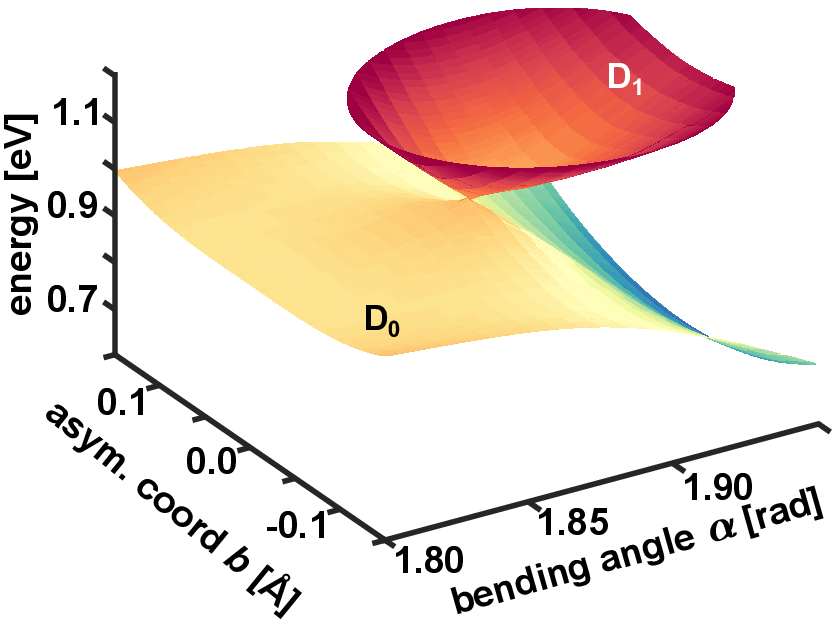}
        \label{fig:CoIn}
    }
    \subfigure[][]{
        \includegraphics[width=0.45\textwidth]{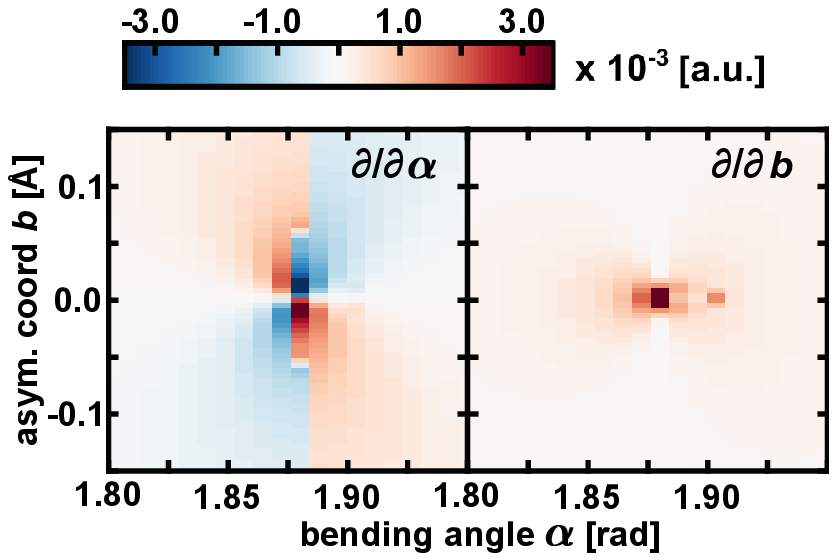}
        \label{fig:Nacs}
    }    
    \caption{\label{fig:PES_CoIn}(a) Adiabatic potential energy surfaces of the $D_0$ state (left) and $D_1$ state (right) of \ce{NO2}. The CoIn is marked in white and the positions of the relevant minima in the two-dimensional subspace are displayed in black.The two marked minima are only slightly higher in energy than the fully optimized minimum structures shown in the SI. (b) The vicinity of the $D_1$/$D_0$-CoIn. (c) Non-adiabatic coupling elements between $D_1$ ans $D_0$ at the CoIn, $\alpha$-element left and $b$-element right.}
\end{figure}

We apply our extended NEMol approach to the non-adiabatic dynamics of \ce{NO2}. In this molecule, a CoIn (depicted in FIG.~\ref{fig:CoIn}) between the $D_1$ and the $D_0$ state enables a radiationless relaxation. The ultrafast non-adiabatic transition takes less than \SI{100}{\femto\second} and has been widely explored both theoretically\cite{Haller:1985,Mahapatra:1999,Santoro:1999,Mahapatra:2000,Santoro:2000,Kurkal:2003,Sanrey:2006,Arasaki:2007,Schinke:2008,Arasaki:2010,Richter:2015,Woerner:2018,Richter:2019} and experimentally\cite{Arasaki:2010,Wilkinson:2010,Arasaki:2011,Woerner:2011,Ruf:2012,Woerner:2012}. First we analyze the relaxation itself and next we apply a few-cycle IR laser pulse to control the dynamics in the vicinity of the CoIn, similar to previous studies\cite{Arasaki:2010,Arasaki:2011,Richter:2015,Richter:2019}. With our NEMol ansatz we can study its influence on the motion of the nuclei and the electrons.   

The nuclear dynamics is performed on the two-dimensional adiabatic potential energy surfaces of the $D_1$ and the $D_0$ state shown in FIG.~\ref{fig:PES_CoIn}. The coordinates spanning the PES's are the gradient difference and derivative coupling vectors defining the branching space of the $D_1$/$D_0$-CoIn depicted in FIG.~\ref{fig:CoIn}. These two vectors correspond to the bending angle $\alpha$ and the asymmetric stretching coordinate $b$, defined as half the difference between the two NO distances. The last internal degree of freedom, the symmetric stretch coordinate, is kept constant at the value of the optimized $D_1$/$D_0$-CoIn~(\SI{1.267}{\angstrom}). As shown by Richter et al.~\cite{Richter:2015} the population dynamics obtained within this two-dimensional coordinate space is in very good agreement with the full dimensional simulations by Arasaki et al.~\cite{Arasaki:2010}. We performed our dynamics simulations in the adiabatic representation and the corresponding NAC's between $D_1$ ans $D_0$ are shown in FIG.~\ref{fig:Nacs}. It should be mentioned that in previous studies\cite{Arasaki:2010,Richter:2015,Richter:2019} the simulations were performed in the diabatic representation and therefore small deviations may occur due to the limitation of the grid spacing. Further information about the simulation setup can be found in section~II of the SI.

\begin{figure}[ht!]
    \centering
        \includegraphics[width=0.45\textwidth]{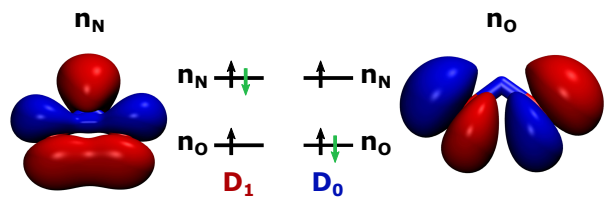}
    \caption{\label{fig:active_orbital}Molecular orbital schema with the active electron indicated in green and corresponding orbitals at the optimized CoIn. Orbitals are shown with an isovalue of $0.05$.}
\end{figure}

In order to calculated the coupled electron density according to equation~\ref{eq:gridDens} we define a NEMol-grid of $15 \times 13$ points which are equally distributed between \SIrange{1.34}{2.86}{\radpi} in the $\alpha$-coordinate and between \SIrange{-0.33}{+0.33}{\angstrom} in the $b$-coordinate. The necessary population- and overlap-terms are calculated for equal-spaced segments around these grid points. To cover the entire PES the segments for the boundary grid points are larger. The transformation of the full wavepacket onto the NEMol-grid, the overlap terms and the resulting coherence terms are visualized in FIG.~S6 (free propagation) and FIG.~S10 (propagation with laser pulse) in the SI. The two active orbitals which are required to describe the NEMol-dynamics in the one-electron-two-orbital (1e-2o) picture are shown in FIG.~\ref{fig:active_orbital} at the optimized CoIn. The non-binding orbital $n_N$ with contributions at the nitrogen atom is associated with the $D_1$ state and the non-binding orbital $n_O$ located only at the oxygen atoms is attributed to the $D_0$ state. The energy difference $\Delta E$ between the $D_0$ and $D_1$ state for each grid point is shown in FIG.~S4 of the SI.

\subsection{\label{subsec:free_dyn} Free dynamics of \ce{NO2}}
\begin{figure*}[ht!]
    \centering
\includegraphics[width=0.90\textwidth]{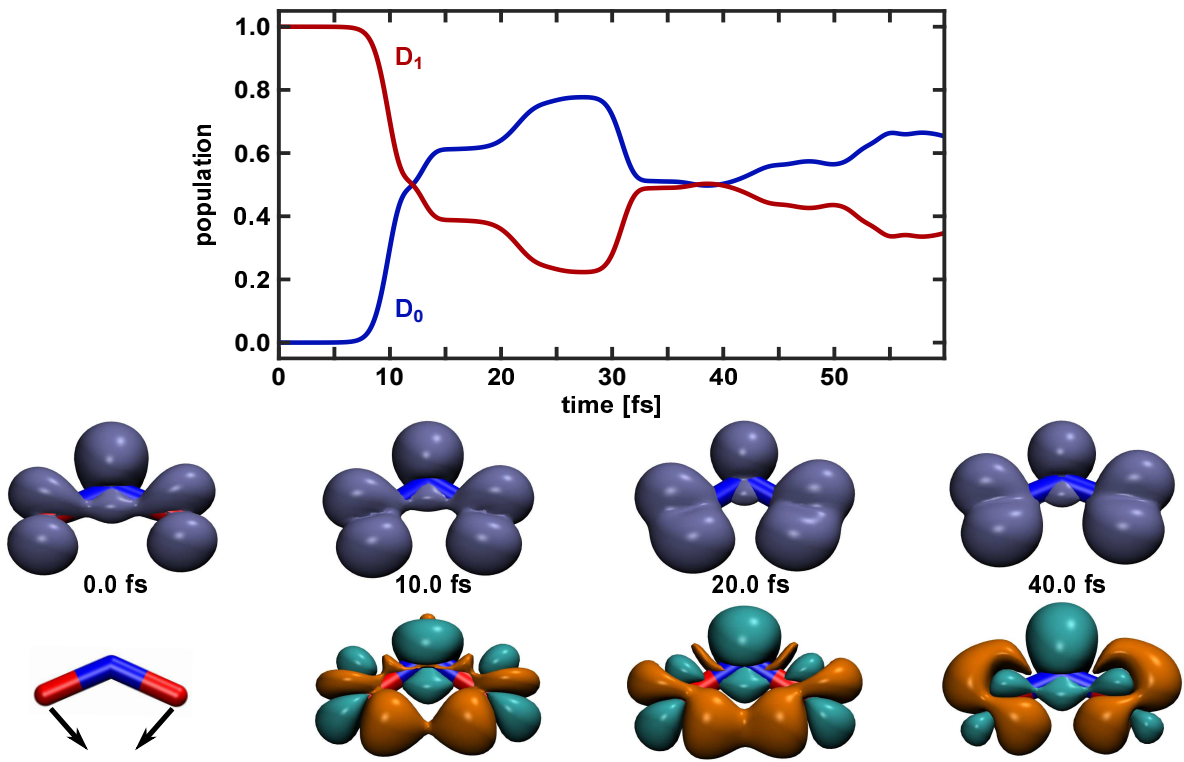}
\caption{\label{fig:pop_dens}Free dynamics of \ce{NO2}. Upper panel: Populations of the $D_0$ and $D_1$ state as a function of simulation time. Lower panel: Snapshots of the electron density in the 1e-2o picture and the difference in density relative to the initial density (green electron-loss, orange electron-gain). The isovalues used are $0.006$ respectively $\pm 0.002$. The arrows on the left geometry (\SI{0.0}{\femto\second}) indicate the movement of the nuclei.}
%An animation of the electron density in the 1e-2o picture and the difference density can be found in the SI.} 
\end{figure*}

To initiate the dynamics simulation in the $D_1$ state we assumed a delta pulse excitation. The temporal evolution of the population of both states is shown in the upper panel of Fig.~\ref{fig:pop_dens} and the dynamics of the nuclear wavepackets integrated over the $\alpha$-coordinate, respectively, the $b$-coordinate are depicted in FIG.~S5 for both surfaces. The nuclear wavepacket started in $D_1$ reaches the vicinity of the CoIn after approximately \SI{7}{\femto\second} for the first time. While passing the coupling region in the time interval from \SIrange{7}{15}{\femto\second} the population of the electronic ground state increases to over \SI{60}{\percent}. The part of the nuclear wavepacket remaining in the $D_1$ state reaches its turning point around \SI{15}{\femto\second} and then propagates backwards. This leads to a second passage through the CoIn area and an increase of the population of the $D_0$ state around \SI{22}{\femto\second}. The nuclear wavepacket evolving on the lower adiabatic surface, re-encounters the CoIn region later at around \SI{30}{\femto\second}. During this third passage, a substantial part of the population is transferred back into the excited state. After \SI{35}{\femto\second} the wavepacket is delocalized on both surfaces and the population is nearly equal in both states. Towards the end of the simulation at around \SI{50}{\femto\second} a fourth passage occurs. The wavepacket remains symmetrical with respect to the $b$-coordinate for the whole simulation time. For the wavepacket on the lower PES (see right sight of \mbox{FIG.~S5(b)} in the SI) the formation of a nodal structure for $b = \SI{0.0}{\angstrom}$ is clearly visible, which is a signature of destructive self-interference due to the geometric phase effect\cite{Berry:1984,Yarkony:1996,Baer:2002}. 

\begin{figure}[ht!]
    \centering
        \includegraphics[width=0.45\textwidth]{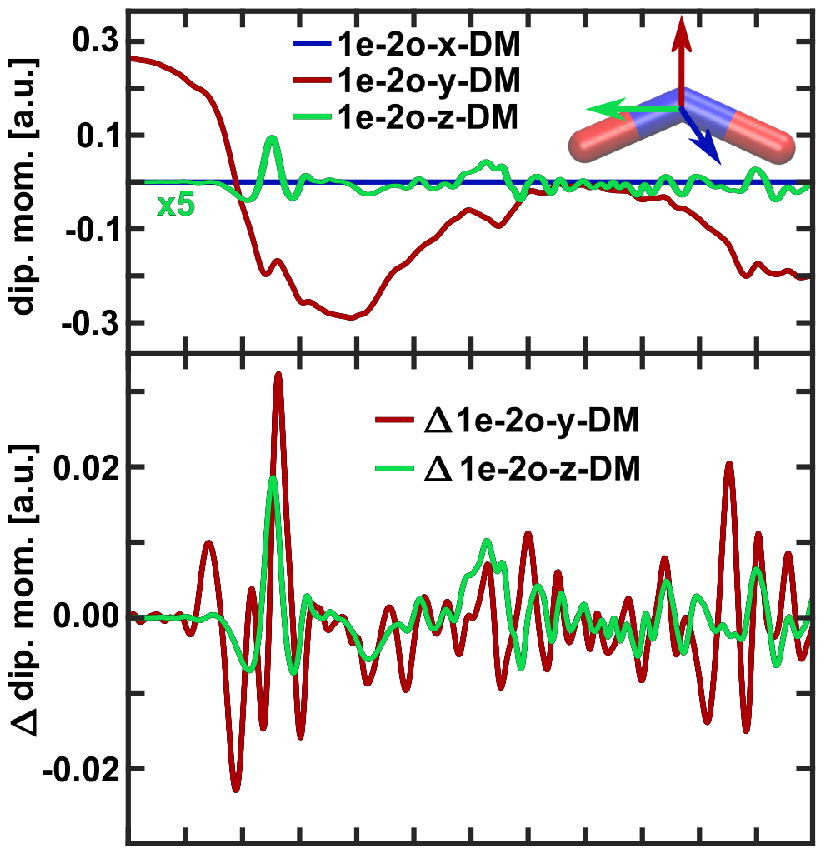}
    \caption{\label{fig:dip_free}Field-free temporal evolution of the dipole moment components based on the electron density in the 1e-2o picture.  Upper panel: total value of all three components. The \mbox{1e-2o-$z$-component} is enhanced by a factor of five. The orientation of the molecule is shown as inlay in the upper right corner.  Lower panel: Difference between the dipole moment components (\mbox{$\Delta$ 1e-2o}) one time calculated with the coherence term included and once without it. Differences only shown for the for $y$- and $z$-components.}
\end{figure}

In the lower part of FIG.~\ref{fig:pop_dens} snapshots of the electron density in the 1e-2o picture are shown. For a better visualization also the difference in density with respect to $t = \SI{0}{\femto\second}$ is depicted. The molecule is orientated in such a way that the molecular plane is equivalent to the $yz$-plane and the center of mass defines the origin of the laboratory frame. Therefore the internal $\alpha$-coordinate points to the same direction as the $y$-coordinate and the internal $b$-coordinate is associated with the $z$-coordinate. The orientation of the molecule is shown in the upper right corner of FIG.~\ref{fig:dip_free}. In correspondence to the non-adiabatic transition from the $D_1$ state to the $D_0$ state, the main feature of the electron dynamics is the loss of density at the nitrogen and the corresponding gain of density at the oxygen atoms. In addition, the change in the electron density attributed to the motion of the nuclei (Born-Oppenheimer part) is present. Due to the high symmetry of \ce{NO2}, the electron density is mirror-symmetrical with respect to the $xy$-plane, which is equivalent to the symmetric behaviour of the nuclear wavepacket with respect to the $b$-coordinate.

To analyze the electron dynamics we calculated the dipole moment of the electron density within the 1e-2o picture. In the upper panel of FIG.~\ref{fig:dip_free} the temporal evolution of its three components is shown; for the molecular orientation see the upper right corner of FIG.~\ref{fig:dip_free}. To distinguish the Born-Oppenheimer part of the dynamics from the coherent electron dynamics the density was calculated once with the coherent part included and once without. For both quantities the respective dipole moments were determined as well as their difference, hereinafter labeled as \mbox{$\Delta$ 1e-2o} and shown in the lower panel of FIG.~\ref{fig:dip_free}. 
The active orbitals do not change along the $x$-coordinate and thus the \mbox{1e-2o-$x$}-component of the dipole moment stays zero and is excluded from further discussions. The \mbox{1e-2o-$y$}-component shows the largest values and the strongest changes over time. Its evolution follows the dynamics of the population. In the initial \SI{20}{\femto\second} the first passage through the CoIn region occurs and simultaneously the value of the \mbox{1e-2o-$y$}-component changes from \SI{+0.3}{\au} to \SI{-0.3}{\au}. The zero crossing occurs at \SI{10}{\femto\second}. For later times when dephasing and partial recurrence of the nuclear wavepackets become important the $y$-component approaches zero at about \SI{40}{\femto\second} and becomes negative thereafter again. These main features disappear for the \mbox{$\Delta$ 1e-2o-$y$}-component (lower panel FIG.~\ref{fig:dip_free}) and only fast oscillations with one order of magnitude smaller amplitudes are left. The largest amplitudes are observed around \SI{10}{\femto\second}, \SI{30}{\femto\second} and \SI{50}{\femto\second}. These amplitudes coincide with the passages of the wavepacket through the CoIn region. The large difference between the \mbox{1e-2o} and the \mbox{$\Delta$ 1e-2o} values means that the dynamics of the $y$-component is dominated by the nuclear motion. That is understandable, since the $y$-coordinate is aligned along the main direction of dynamics ($\alpha$-coordinate), which mediates the non-adiabatic transition. The temporal evolution of the \mbox{1e-2o-$z$}-component is an order of magnitude smaller and almost identical to its $\Delta$ value. The dynamics of the $z$-component is not dominated by the nuclear motion but solely induced by the coherent electron dynamics. Therefore, we can use the $y$ and the $z$ component to distinguish between the two contributions of the coupled electron dynamics. As the $\Delta$ values of both components lie amplitude wise in the same region and show a similar pattern they are suitable to monitor the coherent electron dynamics in the system.  Overall the nuclear motion has a much larger impact on the dipole moment than the coherent electron dynamics.

\begin{figure}[ht!]
    \centering
    \subfigure[][]{
        \includegraphics[width=0.45\textwidth]{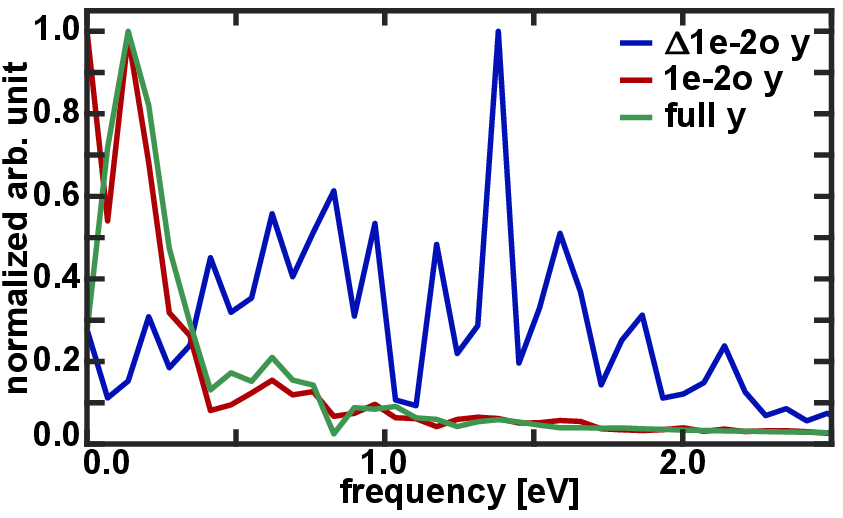}
        \label{fig:spectra_y_free}
    }
    \subfigure[][]{
        \includegraphics[width=0.45\textwidth]{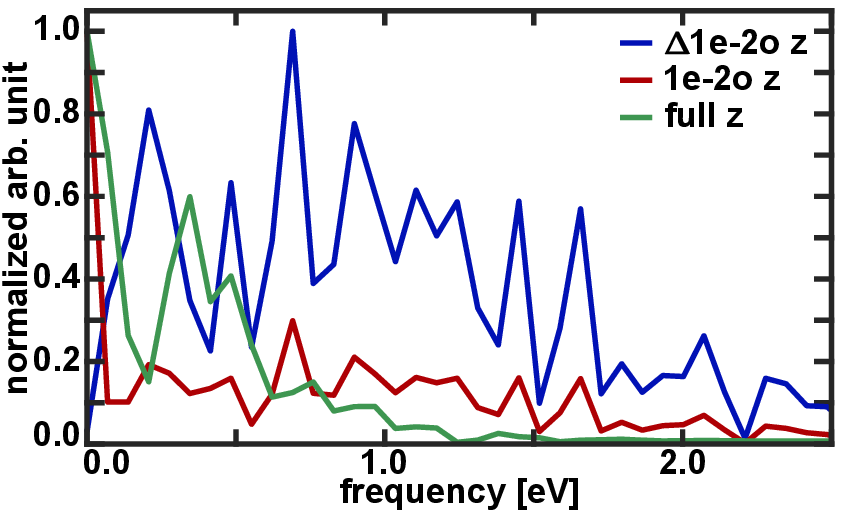}
        \label{fig:spectra_z_free}
    }
    \caption{\label{fig:spectra_yz_free} The Fourier spectra of the $y$-component (a) and $z$-component (b) of the dipole moment obtained using the $\Delta$ 1e-2o-components (blue), the 1e-2o-components (red) as well as the components calculated with the full density (green). All spectra are normalized to one individually.}
\end{figure}

By applying the Fourier transform to the temporal evolution of the dipole moments the corresponding frequencies are determined. Beside the \mbox{$\Delta$ 1e-2o}- and the 1e-2o-components also the dipole moment calculated with the full density was used. The resulting spectra for the $y$- and the $z$-component for all three cases are shown in FIG.~\ref{fig:spectra_yz_free}. The spectra are all normalized to one individually. The relative magnitude between all quantities can be estimated from figure FIG.~\ref{fig:dip_free}. All frequencies with an intensity larger than 0.1 are listed in \mbox{TABLE~S3} and \mbox{TABLE~S4} of the SI. 

 The \mbox{$\Delta$ 1e-2o} spectra (FIG.~\ref{fig:spectra_yz_free} blue), reflecting the coherent electron dynamics, cover the largest frequency range from \SIrange{0.2}{2.3}{\electronvolt} for both components, whereas the energy differences $\Delta E$ (\SIrange{0.0}{1.0}{\electronvolt} see FIG.~S4) in the vicinity of the CoIn,  which enter in the coherent part of the electronic wavepacket, are smaller. These discrepancy can be rationalized when taking a closer look at the coherence term (see equation~\ref{eq:coupledDens}). Two of the factors in the product contribute to the overall phase, the nuclear overlap and the electronic phase term containing the $\Delta E$ values. The phase of the overlap term relates to the difference in momentum of the nuclear wavepackets involved. In our test system \ce{NO2} the wavepacket on $D_1$ approaches the CoIn with a high momentum, larger than the $\Delta E$ gaps near the CoIn. In other words the coherent dynamics of the electronic wavepacket is in the \ce{NO2} case also significantly influenced by the phase-differences of the nuclear wavepackets moving on different potentials. This correlation is illustrated in in FIG.~S7 in the SI for two individual NEMol-grid points.
The frequencies for the 1e-2o-components (FIG.~\ref{fig:spectra_yz_free} red) are dominated by the slower nuclear dynamics (Born-Oppenheimer part) giving rise to the strong peaks below \SI{0.5}{\electronvolt}. Simultaneously, high energy parts lose intensity. This effect is stronger for the $y$-component, whereas for the $z$-component the initial pattern is still recognizable. This behaviour is further increased for the full density (FIG.~\ref{fig:spectra_yz_free} green). For both components some peaks appear in all three cases, especially in the energy range between \SIrange{0.5}{0.75}{\electronvolt}. They can be attributed to the coherent electronic dynamics and may also be experimentally observable.

\begin{figure*}[ht!]
\includegraphics[width=0.90\textwidth]{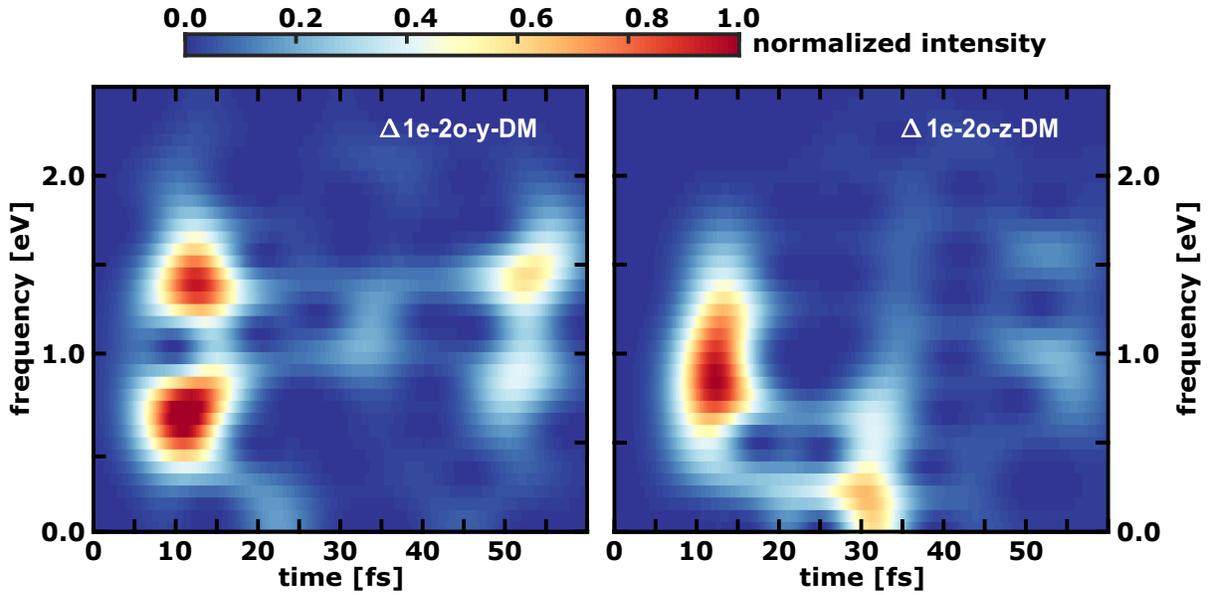}
\caption{\label{fig:dip_free_2d}Short-time Fourier transform of the \mbox{$\Delta$ 1e-2o}-$y$ dipole moment component (left) and \mbox{$\Delta$ 1e-2o}-$z$ dipole moment component (right). The Fourier spectrograms are normalized and a Gaussian windowing function with a width of 180 data points corresponding to a time of \SI{18.14}{\femto\second} is used.}
\end{figure*}

Further information can be gained by extracting the time when these frequencies occur. This allows us to connect them to a specific movement in the system. Therefore, we performed short-time Fourier transform spectra for the \mbox{$\Delta$ 1e-2o}-$y$ and the \mbox{$\Delta$ 1e-2o}-$z$ components using a Gaussian windowing function with a width of 180 data points corresponding to a time of \SI{18.14}{\femto\second}. The resulting two spectrograms are shown in FIG.~\ref{fig:dip_free_2d}.
The \mbox{$\Delta$ 1e-2o}-$y$ spectrogram (left) shows two main pairs of signals around \SI{10}{\femto\second} (\SIrange{0.5}{1.7}{\electronvolt}) and \SI{50}{\femto\second} (\SIrange{0.7}{1.7}{\electronvolt}, which correspond to the first and the fourth passage of the wavepacket through the CoIn region. The signals are most pronounced at the first passage and significantly attenuated at the fourth passage. There are considerably weaker peaks observable at \SI{25}{\femto\second} and \SI{30}{\femto\second}, which can be attributed to the second and the third passage. Also the \mbox{$\Delta$ 1e-2o}-$z$ spectrogram (right) shows two main signals. The first one appears around \SI{10}{\femto\second} (first passage through CoIn) and covers a frequency range from \SIrange{0.5}{1.7}{\electronvolt}. The third passage around \SI{30}{\femto\second} can be attributed to the second signal which extends over low-frequency components (\SIrange{0.1}{1.0}{\electronvolt} and has a lower intensity. Again considerably weaker peaks can be found around \SIrange{20}{25}{\femto\second} (second passage) and after \SI{50}{\femto\second} (fourth passage).
Thus each passage of the nuclear wavepacket through the CoIn region induces coherent electron dynamics, although not to the same extent for both components. The coherent dynamics is only short-lived for \SIrange{5}{7}{\femto\second} and the intensity of its signal decreases with time. The highest intensities are observed for the first transition when the localized initial nuclear wavepacket hits the CoIn. The subsequent dephasing and branching of the nuclear wavepacket blurs the electronic coherence. In summary, we observe a short but recurring appearance of the coherent electron dynamics that is modulated by the nuclear wavepacket motion. In the following we focus on the first passage (\SI{10}{\femto\second}) for applying a few-cycle IR pulse to influence the coupled dynamics of \ce{NO2}, since here the largest electronic coherence in the field-free case exists.

\subsection{\label{subsec:cep_dyn} Dynamics in the presence of a few-cycle IR pulse}
Again a delta pulse excitation is used to initiate the dynamics. With the appropriate time delay, a few-cycle IR laser pulse is applied to influence the first passage through the CoIn and thereby the subsequent coupled dynamics. The used few-cycle pulse has a Gaussian shape and is defined as:
\begin{eqnarray} 
    E(t) & = & E_{max}\cdot e^{-2\left(\frac{t-t_0}{\sigma}\right)^2} \cdot \cos{\left(\omega_0\left(t-t_0\right)+\phi\right)}, \\
    \text{with}~\sigma & = & \frac{\rm\small{FWHM}}{\sqrt{2\log\left(2\right)}}.  \nonumber
\end{eqnarray}
with the central frequency $\omega_0$, the time zero $t_0$, the maximal field amplitude $E_{max}$, the full width half maximum (FWHM) and the carrier envelope phase $\phi$ (CEP). The time zero $t_0$ of the pulse, defining the position of its maximum, was chosen to match the time window when the wavepacket is located near the CoIn ($t_0 = \SI{10}{\femto\second}$). For this time the nuclear wavepacket is still very localized and the electronic coherence maximal. The central frequency $\omega_0$ is chosen to be resonant with the actual energy gap $\Delta E$= \SI{0.76}{\electronvolt} between the electronic states. The remaining three pulse parameters, the field amplitude $E_{max}$, the full width half maximum (FWHM) and the CEP $\phi$, are set to $E_{max} =$~\SI{0.103}{\giga\volt\per\centi\metre}(which corresponds to a maximum intensity of  \SI{1.4e13}{\watt\per\square\centi\metre}), FWHM~$=$~\SI{8}{\femto\second} and $\phi = 0 \pi$. In comparison with the pulse parameters used by Richter~et~al.\cite{Richter:2015,Richter:2019} all values are quite similar. Only our intensity is lower to stay in the range where the influence of the CEP pulse is mainly determined by the interplay of the non-adiabatic transition and the light induced electronic coherence\cite{Schueppel:2020}. By this we also ensure to stay below or at the threshold of ionization. The light-matter interaction is treated within the dipole approximation, for details see section~I in the SI. We assume, that the electric component of the pulse is optimally aligned with the transition dipole moment. The absolute value of the TDM is used, which is shown in FIG.~S3(a) of the SI. As stated by Richter~et~al.\cite{Richter:2015} already a moderate molecular alignment distribution is sufficient to observe the effect of such a control pulse.

\begin{figure*}[ht!]
\includegraphics[width=0.9\textwidth]{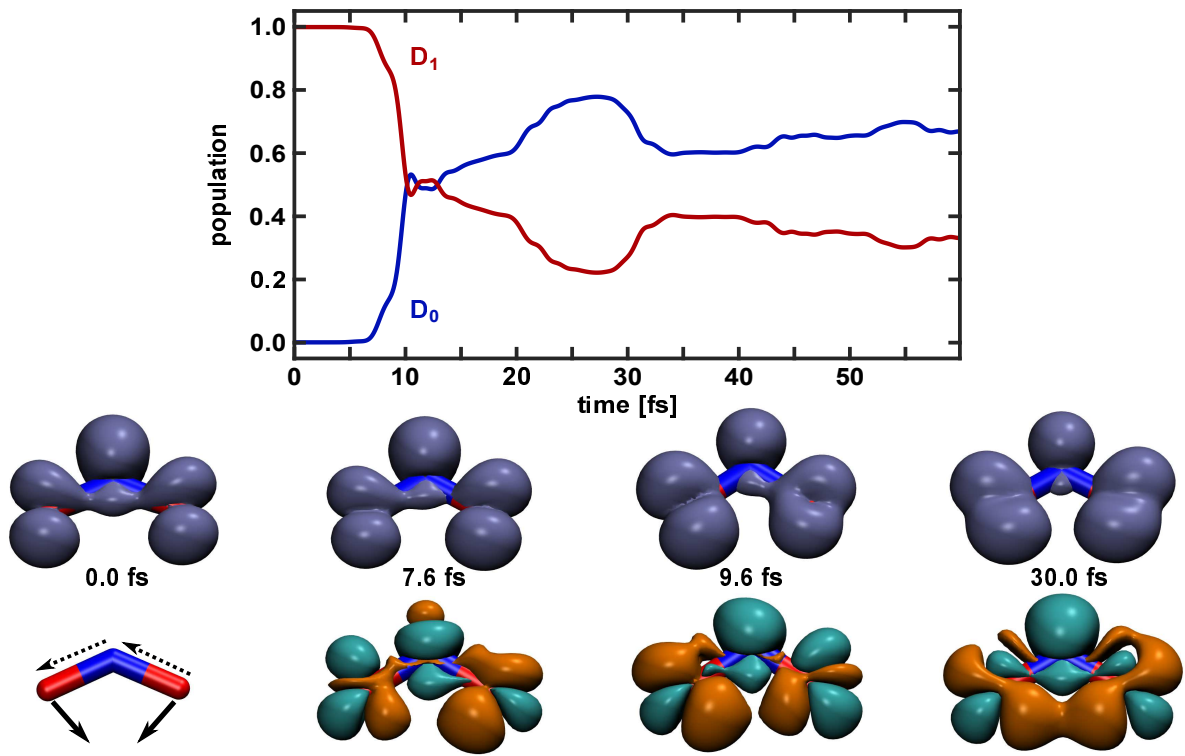}
\caption{\label{fig:pop_dens_cep} Dynamics of \ce{NO2} in the presents of a few-cycle IR laser pulse. Upper panel: Populations of the $D_0$ and $D_1$ state as a function of simulation time. Lower panel: Snapshots of the electron density in the 1e-2o picture and the density difference relative to the initial density (green electron-loss, orange electron-gain). The isovalues used are $0.006$ respectively $\pm 0.002$. The arrows on the left geometry (\SI{0.0}{\femto\second}) indicate the movement of the nuclei (main direction as bold and additional movement as dotted arrows).}
%An animation of the electron density in the 1e-2o picture and the difference density can be found in the SI.} 
\end{figure*}

The evolution of the adiabatic populations influenced by the few-cycle IR-field is shown in the upper panel of FIG.~\ref{fig:pop_dens_cep}. The related nuclear wavepacket dynamics on both surfaces integrated over the $\alpha$-coordinate, respectively, the $b$-coordinate are depicted in FIG.~S8 of the SI.
During the first transition through the CoIn region (\SIrange{7}{15}{\femto\second}) a 50:50 population of both states is created. The interaction with the light pulse is reflected in the small wriggles around \SI{10}{\femto\second}. The subsequent dynamics is comparable to the field-free case up to \SI{30}{\femto\second}. Thereafter no clear passage through the CoIn region is observable. Thus the IR pulse induces a change in the nuclear dynamics which persists beyond the pulse duration. As an important consequence, the nuclear motion becomes asymmetric with respect to the $b$-coordinate and the nuclear wavepacket even loses its nodal structure (compare FIG.~S5(b) and FIG.~S8(b) both in the SI), which was also observed by Richter~et~al.\cite{Richter:2015}. This asymmetry leads to the partly deviations from of the CoIn region after \SI{30}{\femto\second}. On the lower panel of FIG.~\ref{fig:pop_dens_cep} snapshots of the electron density in the \mbox{1e-2o} picture are shown. Again the difference in density with respect to $t = \SI{0}{\femto\second}$ is depicted. The main features in the dynamics are quite similar to the field-free case. But like for the nuclear motion the dynamics of the electron density becomes asymmetric with respect to the $xy$-plane i.e. the $b$-coordinate. This asymmetry persists after the laser pulse is no longer active (for example see the snapshots at \SI{30.0}{\femto\second}). The oscillation of the electron density from the right to the left oxygen is most prominently observable for the snapshots at \SI{7.6}{\femto\second} and \SI{9.6}{\femto\second}.

\begin{figure}[ht!]
    \centering
        \includegraphics[width=0.45\textwidth]{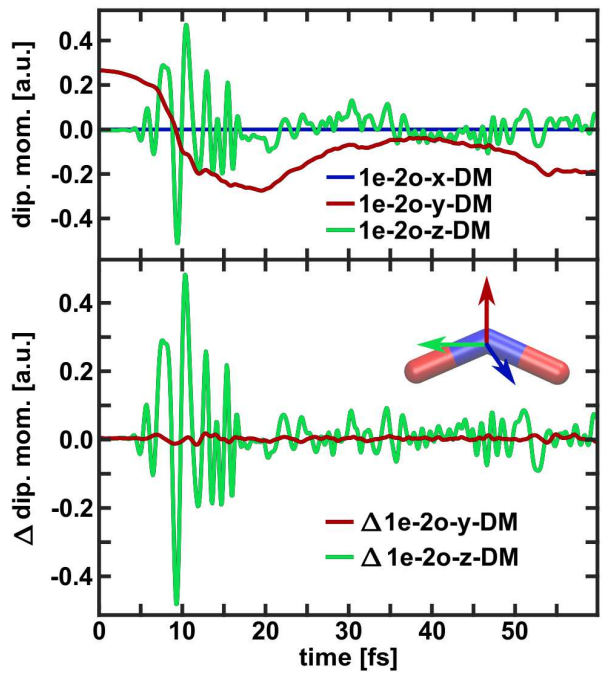}
    \caption{\label{fig:dip_cep}Temporal evolution of the dipole moment components (DM comp) based on the electron density in the 1e-2o picture in the presents of a few-cycle IR pulse. Upper panel: total value of all three components. The orientation of the molecule is shown as inlay in the middle. Lower panel: Difference between the dipole moment components one time calculated with the coherence term included and once without it. Differences only shown for the for $y$- and $z$-DM comp.}
\end{figure}

The oscillations of the electron density are again recorded by the three dipole moment components, shown in the upper panel of FIG.~\ref{fig:dip_cep}. The coherent part of electron dynamics is visualized by the \mbox{$\Delta$ 1e-2o} dipole moment components for the $y$- and $z$-coordinate in the lower panel.
Again the 1e-2o-$x$-component stays zero for the whole simulation time. As the few-cycle IR pulse induces the asymmetry mainly along the $b$-coordinate, the overall temporal evolution of the 1e-2o-$y$-component and the \mbox{$\Delta$ 1e-2o}-$y$ is similar to the field-free case. The 1e-2o-$z$-component experiences the main changes. During the pulse strong and fast oscillations are observed with an amplitude nearly thirty times larger than for the field-free case. The oscillations stay up to ten times larger after the pulse. The superimposed slow oscillation with a period of about \SI{20}{\femto\second} can be assigned to the asymmetry in the nuclear motion. It does not appear for the \mbox{$\Delta$ 1e-2o}-$z$ component reflecting solely the coherent electron dynamics. By breaking the symmetry of the nuclear motion with the laser pulse the electronic coherence induced in the \ce{NO2} molecule is significantly larger. Again it is observable mainly in the $z$-component, respectively, in the $b$-coordinate. During the light pulse it is now the coherent electron dynamics which is responsible for the largest changes in the dipole moment.

\begin{figure}[ht!]
    \centering
    \subfigure[][]{
        \includegraphics[width=0.45\textwidth]{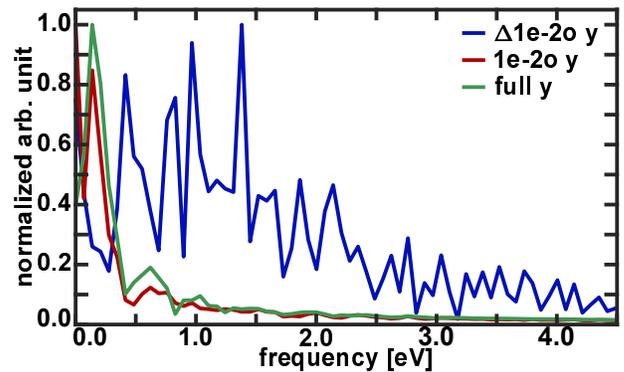}
        \label{fig:spectra_y_cep}        
    }
    \subfigure[][]{
        \includegraphics[width=0.45\textwidth]{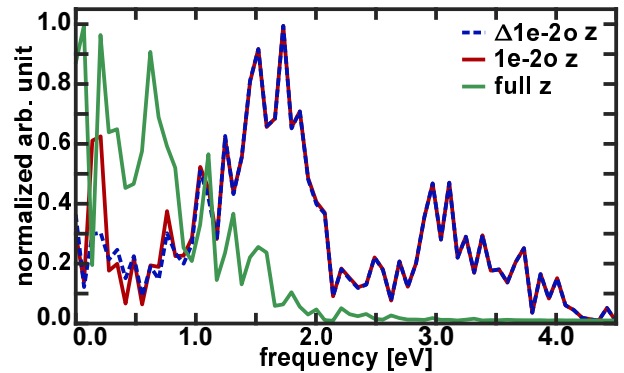}
        \label{fig:spectra_z_cep}            
    }
    \caption{\label{fig:spectra_yz_cep} The Fourier spectra of the $y$-component (a) and $z$-component (b) of the dipole moment in the presents of a CEP-pulse ($0.0 \pi$) obtained using the $\Delta$ 1e-2o-components (blue), the 1e-2o-components (red) as well as the components calculated with the full density (green). All spectra are normalized to one individually.}
\end{figure}

The corresponding frequencies for the \mbox{$\Delta$ 1e-2o}-, the \mbox{1e-2o}-components as well as for the dipole moment calculated with the full density are again determined by Fourier transform. Their spectra are shown in FIG.~\ref{fig:spectra_yz_cep}. All frequencies  with an intensity larger than 0.1 are listed in \mbox{TABLE~S5} and \mbox{TABLE~S6} of the SI.
In both \mbox{$\Delta$ 1e-2o} spectra frequencies up to \SI{4.0}{\electronvolt} appear, which are higher compared to the field-free case. As expected, the main peaks of the \mbox{$\Delta$ 1e-2o-$y$} spectra (FIG.~\ref{fig:spectra_y_cep} blue) are in the same energy region as in the field-free case and only the \mbox{$\Delta$ 1e-2o} $z$-spectrum (FIG.~\ref{fig:spectra_z_cep} blue dotted line) shows differences. Its main peaks are shifted to higher energies by roughly \SI{0.7}{\electronvolt}. The laser pulse injects energy (\SI{0.76}{\electronvolt}) into the system, which influences the momentum of the nuclear wavepacket and thereby the phase of the overlap term (equation~\ref{eq:coupledDens}) which subsequently leads to higher frequencies observed in the coherent electron dynamics. The correlation between the phase of the overlap term, the electronic phase and the laser pulse is illustrated in FIG.~S11 of the SI for two individual grid points.
The frequencies for the $y$-component determined with the 1e-2o-density (FIG.~\ref{fig:spectra_y_cep} red) and the full-density (FIG.~\ref{fig:spectra_y_cep} green) exhibit the same behaviour as in the field-free case. The high energy parts lose significantly intensity since the slower nuclear dynamics (Born-Oppenheimer part) dominates this signal. The dominance of the oscillating dipole moment originating from the coherent electron dynamics shows up in the nearly identical spectra for the 1e-2o-$z$ (FIG.~\ref{fig:spectra_z_cep} red) and \mbox{$\Delta$ 1e-2o}-$z$ (FIG.~\ref{fig:spectra_z_cep} blue). For the $z$-spectra of the full-density (FIG.~\ref{fig:spectra_z_cep} green)  the high energy parts lose some intensity but still more high energy contributions survive compared to the field-free case. 

\begin{figure*}
\includegraphics[width=0.90\textwidth]{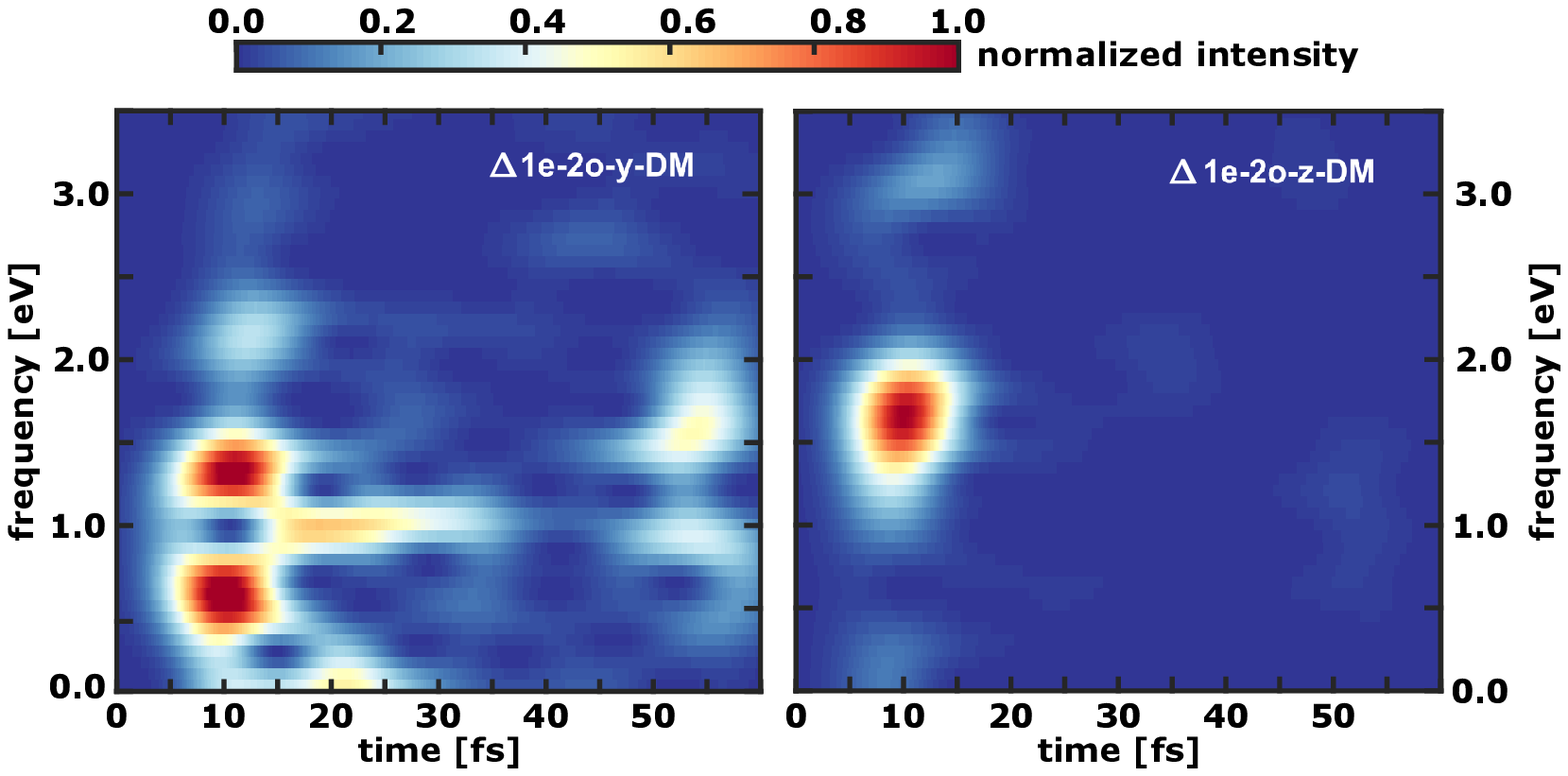}
\caption{\label{fig:dip_cep_2d}Short-time Fourier transform of the \mbox{$\Delta$ 1e-2o}-$y$ dipole moment component (left) and \mbox{$\Delta$ 1e-2o}-$z$ dipole moment component (right) with a few-cycle IR pulse included in the simulation. The Fourier spectrogramms are normalized and a Gaussian windowing function with a width of 180 data points corresponding to a time of \SI{18.14}{\femto\second} is used.}
\end{figure*}

The results of the short-time Fourier transform for the \mbox{$\Delta$ 1e-2o}-$y$ and the \mbox{$\Delta$ 1e-2o}-$z$ dipole moment component using a Gaussian windowing function with a width of 180 data points corresponding to a time of \SI{18.14}{\femto\second} are shown in FIG.~\ref{fig:dip_cep_2d}.
Both spectrograms show a dominant signal which is attributed to the first passage through the CoIn region. The observable electron dynamics is significantly strengthened by the simultaneous light pulse interaction. In case of the \mbox{$\Delta$ 1e-2o}-$y$ spectrogram (left) some new features between \SIrange{10}{30}{\femto\second} appear. Due to the symmetry breaking of the nuclear motion by the laser pulse, signals with very low frequencies as well as an extended signal around \SI{1.0}{\electronvolt} appear. For the more affected \mbox{$\Delta$ 1e-2o}-$z$ component only one dominant peak is observed. 
In summary, the presence of a few-cycle IR pulse modifies the coupled dynamics by breaking the symmetry of the nuclear motion and changing the temporal evolution of the population. Both factors lead to a significant increase of electronic coherence in the molecule especially along the $z$-coordinate (laboratory frame), respectively,  the $b$-coordinate (internal frame).

\section{\label{sec:CEP}Waveform control of molecular dynamics}
In the last part we investigate the controllability of  the nuclear and electron dynamics by the variation of the CEP $\phi$ of a few-cycle IR laser pulse. As shown in the literature\cite{Weitzel:2007,Kling:2008,Znakovskaya:2011,Znakovskaya:2012,Alnaser:2014,Arnold:2018,Schueppel:2020} the CEP control scheme offers the possibility to steer electrons and nuclei in the ionization process but also during the passage through a CoIn. The few-cycle IR pulse builds up a coherent electronic and nuclear wavepacket with a well-defined phase-relationship controllable by the CEP. In the vicinity of a CoIn also the non-trivial geometric phase (Pancharatnam–Berry phase) is introduced\cite{Longuet-Higgins:1958,Berry:1984,Yarkony:1996,Baer:2002}. The interplay of both phase-terms lead to an interference process when the CoIn is passed. The interference (constructive or destructive) can be manipulated by the CEP.

\subsection{\label{subsec:nuc_control} Control of the nuclear dynamics}
\begin{figure}[ht!]
    \centering
        \includegraphics[width=0.45\textwidth]{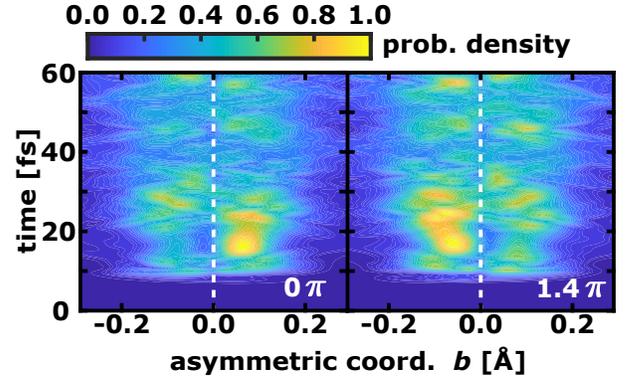}
    \caption{\label{fig:asym_nuc_cep} Normalized nuclear probability density evolution in the presents of a few-cycle IR pulse withe a CEP of $0.0 \pi$ (left) and $1.4 \pi$ (right) on the adiabatic $D_0$-surface integrated over the $\alpha$-coordinate. For the other probability densities see FIG.~S8 and FIG.~S9 in the SI.}
\end{figure}

As a first step we focus on the controllability of the nuclear dynamics. Therefore, we define control objectives which are directly accessible via the nuclear wavepacket and use the population $P_{D0}(t,\phi)$ of the $D_0$ ground state as reference.
\begin{equation}
    P_{D0}(t,\phi) = \bra{\chi_{D0}(R,t,\phi)}\ket{\chi_{D0}(R,t,\phi)}_{R}.
\end{equation}
One objective is the CEP efficiency $\Gamma(t)$ \cite{Schueppel:2020} which is calculated as the difference of the maximum and the minimum population $ P_{D0}(t,\phi)$ for each time step:
\begin{equation}
    \Gamma(t) = \text{max} \left( P_{D0}(t,\phi) \right)  - \text{min} \left( P_{D0}(t,\phi') \right).
\end{equation}
For its maximum value the population of the target state shows the highest CEP-dependence and consequently the highest degree of controllability with respect to the population transfer. The light pulse amplifies the coherent electron dynamics in the system by breaking the symmetry with respect to the asymmetric stretching coordinate $b$, as shown in section~\ref{subsec:cep_dyn}. Therefore, the second objective is the CEP-dependent asymmetry parameter $AN(t,\phi)$ quantifying the CEP induced asymmetry in the nuclear motion with respect to the coordinate $b$.
\begin{equation}
    AN(t,\phi) = \frac{ P_{D0}^R(t,\phi) -  P_{D0}^L(t,\phi)}{P_{D0}(t,\phi)}.
\end{equation}
Where $P_{D0}^L(t,\phi)$ and $P_{D0}^R(t,\phi)$ are defined as follows:
\begin{eqnarray}
P_{D0}^L(t,\phi) & = &  \int\limits_{\alpha_{min}}^{\alpha_{max}} d\alpha \int\limits_{b_{min}}^{0} db \chi_{D0}^{*}(R,t,\phi) \chi_{D0}(R,t,\phi).\\
P_{D0}^R(t,\phi) & = &  \int\limits_{\alpha_{min}}^{\alpha_{max}} d\alpha \int\limits_{0}^{b_{max}} db \chi_{D0}^{*}(R,t,\phi) \chi_{D0}(R,t,\phi).
\end{eqnarray}
In the spirit of the efficiency $\Gamma(t)$ a maximal asymmetry $AN_{max}(t)$ is calculated as:
\begin{equation}
    AN_{max}(t) = \text{max} \left( AN(t,\phi)\right)  + \text{min} \left( AN(t,\phi') \right).
\end{equation}
For its maximum the motion of the nuclear wavepacket shows the highest asymmetry and controllability. Its CEP dependence is illustrated in FIG.~\ref{fig:asym_nuc_cep}.

\begin{figure}[ht!]
    \subfigure[][]
        {\includegraphics[width=0.45\textwidth]{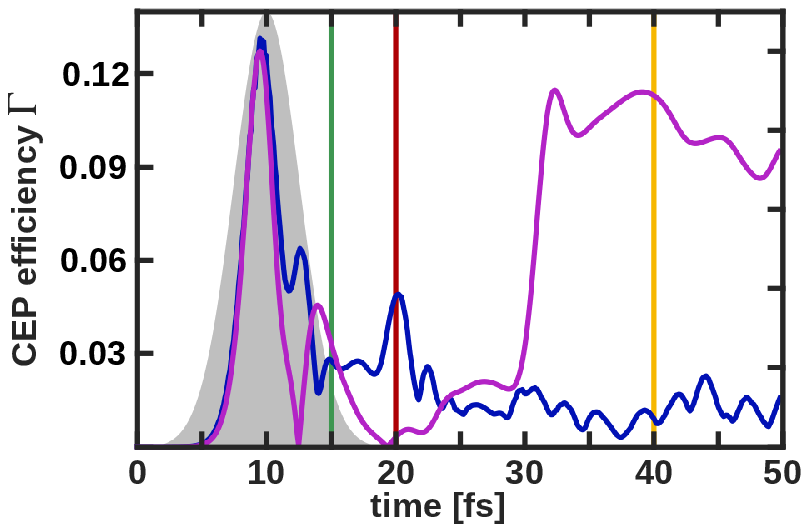}
        \label{fig:eff_pop_abs}}
    \subfigure[][]
        {\includegraphics[width=0.45\textwidth]{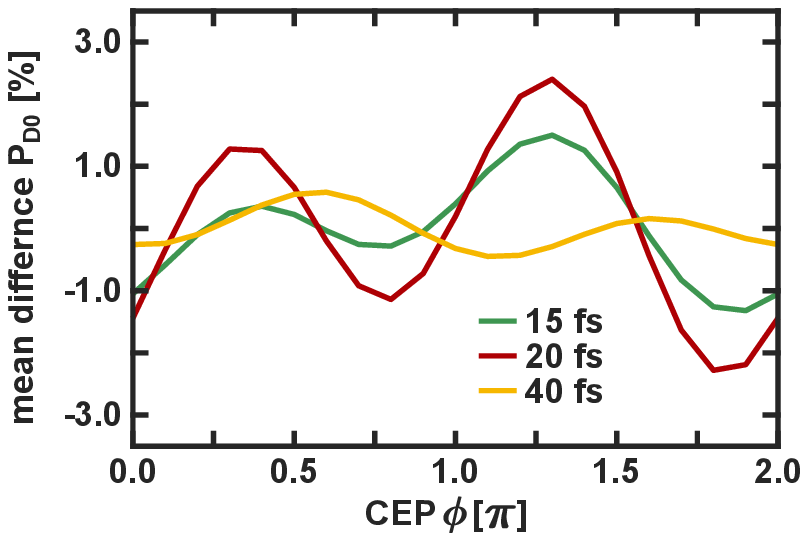}
        \label{fig:pop_abs}}
    \caption{\label{fig:abs_tdm_pop}(a) Temporal evolution of the CEP efficiency $\Gamma(t)$ (blue). The vertical colored lines indicate the points in time that are examined more closely. The violet curve indicates the deviation of the mean population (averaged over all CEP's) from the population in the field-free case. The envelope of the IR pulse is indicated in grey. (b) Mean difference of the CEP-dependent populations $P_{D0}(t,\phi)$ given in percent for different times.}
\end{figure}

The temporal evolution of $\Gamma(t)$ and the CEP dependent population $P_{D0}(t,\phi)$ at three selected times are shown in FIG.~\ref{fig:abs_tdm_pop}.
The CEP efficiency (blue line) reaches its global maximum (\SI{13}{\percent}) nearly simultaneously with the peak intensity ($t_0 = \SI{10}{\femto\second}$) of the laser pulse (grey area). The increase of $\Gamma(t)$ is slightly delayed and the subsequent decrease to \SI{3}{\percent} occurs in two steps. After the laser pulse, approximately at \SI{15}{\femto\second}, $\Gamma(t)$ has a finite oscillating value with a maximum of about \SI{5}{\percent} around \SI{20}{\femto\second}, which indicates the second passage through the CoIn region. The later passages through the CoIn region at \SI{30}{\femto\second} and after \SI{40}{\femto\second} can roughly be seen in the increase of $\Gamma(t)$. The deviation (violet curve) of the mean population (averaged over all CEP's) from the population in the field-free case is significant, especially during the IR pulse and after \SI{30}{\femto\second}.  As discussed with respect to FIG.~\ref{fig:pop_dens_cep}, the induced asymmetry leads to a partial missing of the CoIn region after \SI{30}{\femto\second}, which is almost independent of the CEP chosen.
The CEP-dependence of the population $P_{D0}(t,\phi)$ (see FIG.~\ref{fig:pop_abs}) is recorded for three selected times marked as vertical lines in~\ref{fig:eff_pop_abs}. For better visualization the mean difference is used here and, unless otherwise stated, in all following respective figures. The first line at \SI{15}{\femto\second} (green) matches the end of the laser pulse. The second (red line) and the third point (yellow line) correspond to the second and fourth passage through the CoIn region. For all three times $P_{D0}(t,\phi)$ shows a sinusoidal oscillation with a periodicity of approximately $\pi$. For interference a periodicity of $2\pi$ should emerge. Thus the observed $\pi$ dependence of the population is an indication that it is due mostly to the temporal asymmetry of the few-cycle laser pulse.\cite{Roudnev:2007,Schueppel:2020}

\begin{figure}[ht!]
    \subfigure[][]
        {\includegraphics[width=0.45\textwidth]{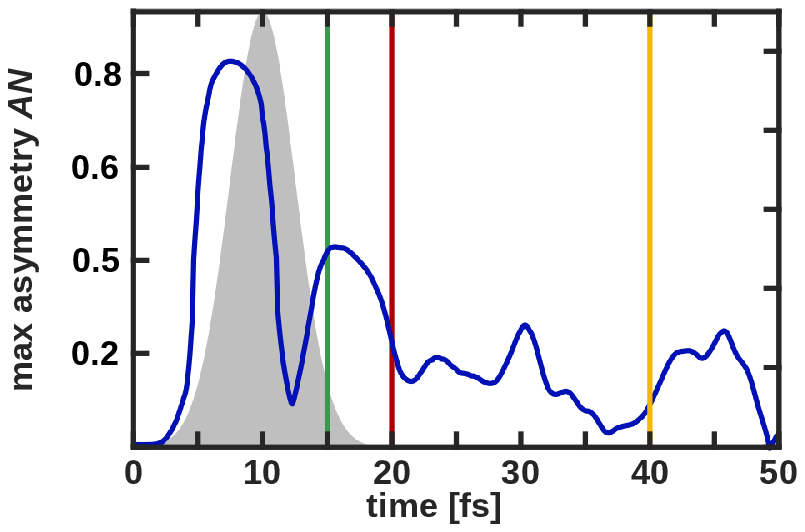}
        \label{fig:eff_asym_abs}}
    \subfigure[][]
        {\includegraphics[width=0.45\textwidth]{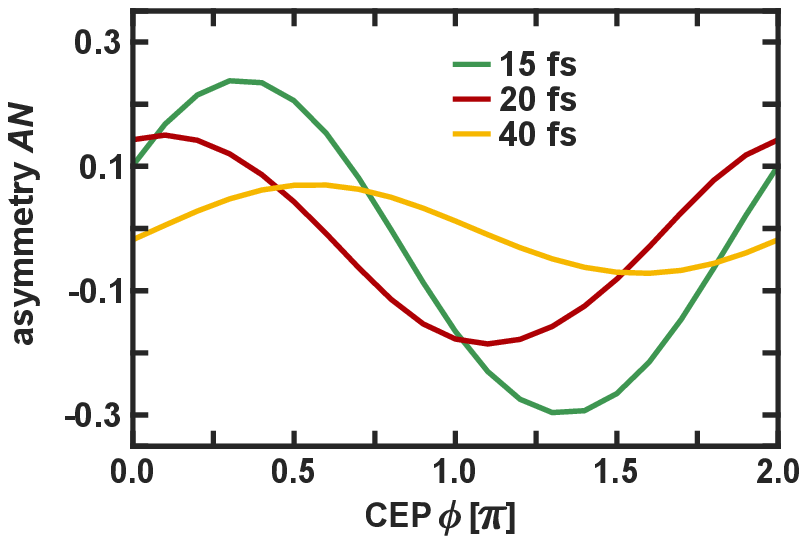}
        \label{fig:asym_abs}}
    \caption{\label{fig:abs_tdm_asym}(a) Temporal evolution of the maximal asymmetry of the nuclei $AN_{max}(t)$ after $t_0 = \SI{10}{\femto\second}$. The vertical colored lines indicate the points in time that are examined more closely. The envelope of the IR pulse is indicated in grey. (b) The CEP-dependent asymmetry parameter $AN(t,\phi)$ for different points in time.}
\end{figure}

An analog analysis is performed for the asymmetry of the nuclear motion along the stretching coordinate $b$ and shown in FIG.~\ref{fig:abs_tdm_asym}.
The maximal asymmetry $AN_{max}(t)$ shows its global maximum around \SI{8}{\femto\second}. As it is defined with respect to the population in $D_0$ alone, the values for the early times (in the beginning of the laser pulse) are overestimated compared to the actual population in the $D_O$ state. Nevertheless, we can deduce that $AN_{max}(t)$ follows the envelope of the laser pulse. The subsequent peaks between \SIrange{15}{20}{\femto\second}, at \SI{30}{\femto\second} and between \SIrange{42}{48}{\femto\second} correspond to the passages through the CoIn region. The decreasing height of the maxima reflects again the delocalization of the nuclear wavepacket with time.
The CEP-dependence of the asymmetry of the nuclear motion (see FIG.~\ref{fig:asym_abs}) $AN(t,\phi)$ is recorded for the same times as previously selected for the CEP-dependent populations $P_{D0}(t,\phi)$. It should be mentioned that the entire value of $AN(t,\phi)$ is shown here and not the mean difference. The asymmetry in the nuclear motion along the coordinate $b$ shows a sinusoidal oscillation, now with a periodicity of $2\pi$ for all three times, which is typical for interference. This means that for the two quantities $P_{D0}(t,\phi)$ and $AN(t,\phi)$ we observe a different CEP-dependence. Or in other words there are two different mechanisms active in the system which can be projected out by using different observables.

In addition we calculated the temporal evolution of $\Gamma(t)$ and $AN_{max}(t)$, as well as the CEP-dependence of  $P_{D0}(t,\phi)$ and  $AN(t,\phi)$  using the $y$-component and the $z$-component of the TDM. Since the results are quite similar the ones obtained with the absolute value of the TDM the orientation of the molecule with respect to electric field of the pulse should not play a major role. For more details see section~IV of the SI.

\subsection{\label{subsec:eleccontrol} Control of the electron dynamics}
\begin{figure}[ht!]
    \subfigure[][]
        {\includegraphics[width=0.45\textwidth]{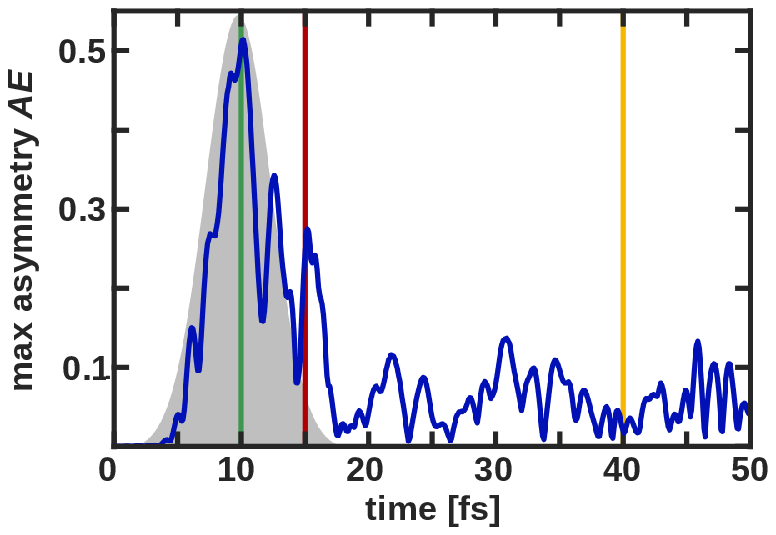}
        \label{fig:asym_max_dens}}
    \subfigure[][]
        {\includegraphics[width=0.45\textwidth]{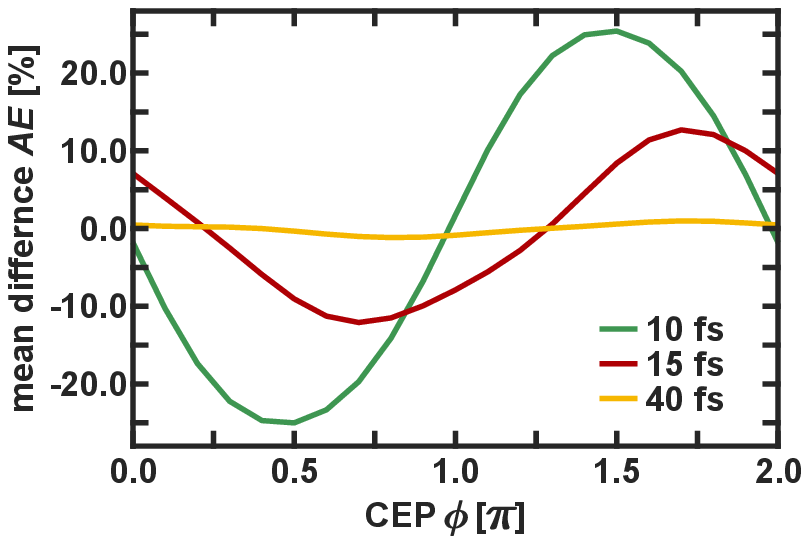}
        \label{fig:asym_dens}}
    \caption{\label{fig:asym_dens_full}(a) Temporal evolution of the maximal asymmetry of the active electron $AE_{max}(t)$. The vertical colored lines indicate the points in time that are examined more closely. The envelope of the IR pulse is indicated in grey. (b) Mean difference of the CEP-dependent asymmetry parameter of the active electron $AE(t,\phi)$ given in percent for different times.}
\end{figure}

As shown in section~\ref{subsec:cep_dyn} the laser pulse is creating a coherent electronic superposition in the vicinity of the CoIn. Therefore, we also examined the influence of the CEP variation on the electron density. The first control objective is the CEP-dependent asymmetry parameter $AE(t,\phi)$ of the \mbox{1e-2o}-density $\rho(r,t,\phi)$.
\begin{equation}
    AE(t,\phi) = \frac{N^R(t,\phi) -  N^L(t,\phi')}{N^R(t,\phi) +  N^L(t,\phi')}.
\end{equation}
with the probabilities $N^L(t,\phi)$ and $N^R(t,\phi)$ to find the electron
on the left or the right side of the molecule given by
\begin{eqnarray}
N^L(t,\phi) & = &  \int\limits_{x_{min}}^{x_{max}} dx \int\limits_{y_{min}}^{y_{max}} dy 
\int\limits_{z_{min}}^{0} dz
 \rho(r,t,\phi).\\
N^R(t,\phi) & = &  \int\limits_{x_{min}}^{x_{max}} dx \int\limits_{y_{min}}^{y_{max}} dy 
\int\limits_{0}^{z_{max}} dz
 \rho(r,t,\phi).
\end{eqnarray}
The maximal asymmetry of the electron density $AE_{max}(t)$ is calculated as follows:
\begin{equation}
    AE_{max}(t) = \text{max} \left( AE(t,\phi)\right)  + \text{min} \left( AE(t,\phi') \right).
\end{equation}
For its maximum the electron dynamics shows the highest CEP-dependence and thus the highest controllability. 
The temporal evolution of $AE_{max}(t)$ and the CEP-dependent asymmetry of the electron density $AE(t,\phi)$ at three selected times are shown in FIG.~\ref{fig:asym_dens_full}.
The maximal asymmetry $AE_{max}(t)$ is highest during the laser pulse (grey area). It decreases within \SI{8}{\femto\second} and becomes smaller by a factor of ten. However during this time period two peaks at \SI{12}{\femto\second} and \SI{15}{\femto\second} can be recognize. Afterwards the maximal asymmetry oscillates between nearly zero and 0.125 until the end of the simulation time. Comparing the maximal asymmetry of the electron density $AE_{max}(t)$ with the one of the nuclei ($AN_{max}(t)$) faster oscillations are observed. To further analyze the response of the electron density (see FIG.~\ref{fig:asym_dens}), $AE(t,\phi)$ is recorded for three selected points in time marked as vertical lines in~\ref{fig:asym_max_dens}). The first line at \SI{10}{\femto\second} (green) corresponds to the main peak of $AE_{max}(t)$ and is taken at the maximum of the pulse. The second point (red line) is taken at \SI{15}{\femto\second} when the laser pulse is approximately over. The last point in time (yellow line) is at \SI{40}{\femto\second}. At all three times $AE(t,\phi)$ shows a sinusoidal oscillation with a periodicity of approximately $2\pi$ and a decreasing amplitude with time. The asymmetry of the electron density thus has the same periodicity as the nuclear asymmetry $AN(t,\phi)$ which is as previously mentioned typical for an interference process.

\begin{figure}[ht!]
    \subfigure[][]
        {\includegraphics[width=0.45\textwidth]{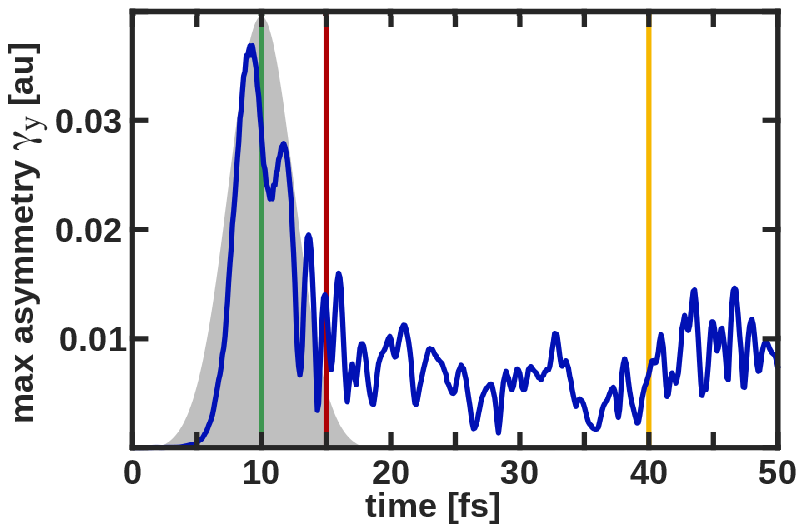}
        \label{fig:asym_max_dipy}}
    \subfigure[][]
        {\includegraphics[width=0.45\textwidth]{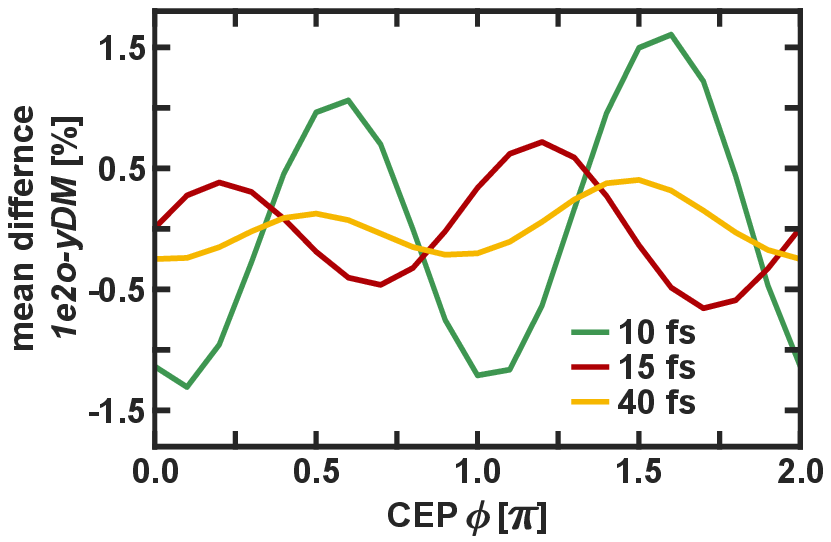}
        \label{fig:asym_dipy}}
    \caption{\label{fig:asym_dipy_full}(a) Temporal evolution of the maximal asymmetry $\gamma_y(t)$ of the \mbox{1e-2o-$y$}-component of the dipole moment. The vertical colored lines indicate the points in time that are examined more closely. The envelope of the IR pulse is indicated in grey. (b) Mean difference of the CEP-dependent \mbox{1e-2o-$y$}-component for different points in time.}
\end{figure}

As already discussed in section~\ref{subsec:cep_dyn} the response of the dipole moment to the applied laser field is an observable directly connected to the electron motion. In the present case the \mbox{1e-2o-$y$}- and the \mbox{1e-2o-$z$}-component are of interest. Their maximal CEP-dependence $\gamma_y(t)$ and $\gamma_z(t)$ are evaluated as the difference of the maximum and the minimum value of \mbox{1e-2o-$y$}-DM$(t,\phi)$ respectively \mbox{1e-2o-$z$}-DM$(t,\phi)$ for each time step. The maximal CEP-dependence $\gamma_y(t)$ is depicted as function of time in FIG.~\ref{fig:asym_max_dipy} and its related component 1e-2o-$y$ in FIG.~\ref{fig:asym_dipy} at three selected times.

The maximal CEP-dependence $\gamma_y(t)$ like all other objectives shows its maximum simultaneously with the maximum of the IR pulse. In this period the shape of the $\gamma_y(t)$ curve is similar to the $\Gamma(t)$ curve (see FIG.~\ref{fig:eff_pop_abs}), only the decrease with decaying pulse intensity is even more asymmetric. After the pulse in the time window from \SIrange{20}{40}{\femto\second} the CEP-dependence oscillates. Again the oscillations are significantly faster than for the nuclear objectives. The CEP-dependence of the \mbox{1e-2o-$y$}-component is recorded in FIG.~\ref{fig:asym_dipy} for the same three selected times as for $AE(t,\phi)$. It shows a sinusoidal oscillation with a periodicity of approximately $\pi$ and a decreasing amplitude with later times. Thus the component shows the same periodicity as $\Gamma(t)$ even with the same phase.

\begin{figure}[ht!]
    \subfigure[][]
        {\includegraphics[width=0.45\textwidth]{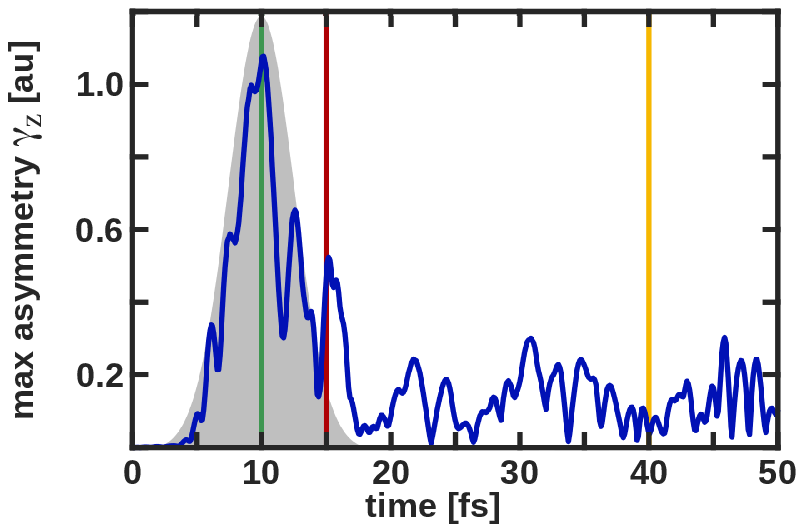}
        \label{fig:asym_max_dipz}}
    \subfigure[][]
        {\includegraphics[width=0.45\textwidth]{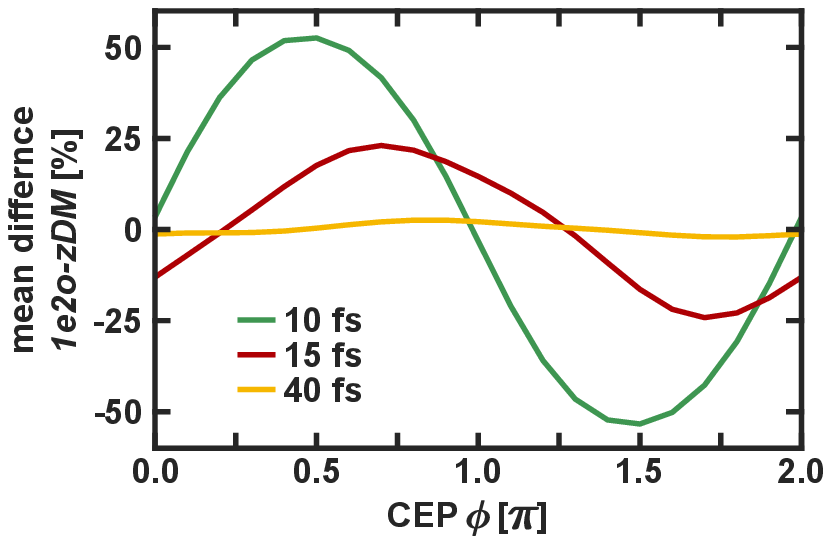}
        \label{fig:asym_dipz}}
    \caption{\label{fig:asym_dipz_full}(a) Temporal evolution of the maximal asymmetry $\gamma_z(t)$ of the \mbox{1e-2o-$z$}-component of the dipole moment. The vertical colored lines indicate the points in time that are examined more closely. The envelope of the IR pulse is indicated in grey. (b) Mean difference of the CEP-dependent \mbox{1e-2o-$z$}-component for different points in time.}
\end{figure}

The temporal evolution of the maximal CEP-dependence $\gamma_z(t)$ and its \mbox{1e-2o-$z$}-component as function of the CEP are shown in FIG.~\ref{fig:asym_dipz_full}.
The maximal CEP-dependence $\gamma_z(t)$ is significantly larger than $\gamma_y(t)$ in consistency with our finding in section~\ref{subsec:cep_dyn} that the $z$-component reacts more strongly to the laser pulse. The overall shape of $\gamma_z(t)$ is quite similar to the temporal evolution of $AE_{max}(t)$ (see FIG.~\ref{fig:asym_max_dens}) and the \mbox{1e-2o-$z$}-component shows the same periodicity of $2\pi$ as $AE(t,\phi)$. The only difference is a phase shift of $\pi$.

In summary, two different responses on the CEP variation are present in the nuclear and electron dynamics. Both asymmetry parameters $AN(t,\phi)$ and $AE(t,\phi)$ as well as the \mbox{1e-2o-$z$}-component of the dipole moment provide a distinction between left and right within the molecular plane ($yz$-plane). The associated $2\pi$ periodicity is typical for an interference process. $\Gamma(t)$ and the \mbox{1e-2o-$y$}-component of the dipole moment are directly sensitive to the main direction of motion along the $\alpha$-coordinate, respectively the $y$-coordinate. The motion in this direction mediates the non-adiabatic transfer between the $D_1$ and $D_O$ state. For these cases the CEP-dependence shows a $\pi$ periodicity, arising from the temporal asymmetry of the few-cycle pulse itself.\cite{Roudnev:2007,Schueppel:2020} Both mechanisms are present for the nuclear as well as for the electron dynamics and can be detected depending on the chosen observable. 

% -----------------------------------------------------------------------------
% CONCLUSION
% -----------------------------------------------------------------------------
\section*{Conclusion}
In this paper, we expand our ansatz for the description of the coupled nuclear and electron dynamics in molecular systems\cite{Geppert:2008,vondenHoff:2009,Znakovskaya:2009} (NEMol). We applied our method to the photoinduced ultrafast dynamics in \ce{NO2} which is dominated by a CoIn. We observe the appearance of a coherent electronic wavepacket at each passage of the CoIn. The coherence is not strong and only short lived due to the high symmetry of the molecule which cancels out the individual contributions\cite{Neville:2020}. Beside the field-free relaxation we also studied the influence of a few-cycle IR laser pulse applied in the vicinity of the CoIn. The induced symmetry breaking significantly enhances the degree of coherence and its life time. Inspired by previous works\cite{Arasaki:2010,Kling:2013,Richter:2015,Arnold:2018,Schueppel:2020} we varied the carrier envelope phase $\phi$ (CEP) of the IR pulse to control the movement of electrons and nuclei during the passage through the CoIn. 

In the first part we generalized our NEMol ansatz. The principle advantage of this ansatz is based on the combination of highly developed quantum-chemical methods with the accurate description of the nuclear quantum dynamics. In the original ansatz\cite{Geppert:2008,vondenHoff:2009,Znakovskaya:2009} an expression for the time-dependent electronic wavepacket is formulated where the electronic part of the total wavefunction is propagated in the electronic eigenstate basis. Its dynamics is extracted from the nuclear wavepacket propagation on coupled potential energy surfaces by introducing the parametric dependence on the time-dependent expected value of  position $\langle R \rangle(t)$. By extending the NEMol ansatz with a grid representation, it is possible to couple the electron dynamics to multiple grid points on which the nuclear wavepacket is represented. Through a simple approximation we were able to condense the coupled dynamics of the one-electron excitation process in the density of one active electron (1e-2o-picture). 
In the second part we compared the coupled nuclear and electron dynamics of \ce{NO2} with and without an IR pulse present when the system reaches the CoIn for the first time. Using the NEMol ansatz, we characterized the coherent electron dynamics by analyzing the temporal evolution of the induced dipole moment. The observed frequencies of the coherent electron dynamics cover a range up to \SI{2.3}{\electronvolt}. These high values originate from the nuclear overlap term as well as from the electronic phase term. In \ce{NO2} the phase contribution of the nuclear overlap term is high and therefor provides a significant contribution to the electron dynamics. The applied few-cycle IR laser pulse generated an asymmetric movement of the nuclear and electronic wavepackets, which is vital for the controllability at the CoIn. The induced oscillating dipole reflects an enhanced build up of the coherent electron dynamics by the laser pulse which survives for several \SI{10}{\femto\second}. 
In the last part the CEP of the IR pulse was varied to influence both the nuclear dynamics as well as the electron dynamics. The CEP-dependent effect lives considerably longer than the pulse in all investigated observables. Depending on the chosen observable a $\pi$ or $2\pi$ periodicity can be found indicating two mechanisms, one based on an interference process ($2\pi$) and the other one induced by the temporal asymmetry of the few-cycle pulse itself ($\pi$). Both periodicities are observed for the nuclear as well as for the electron dynamics. In each case they can be projected out by using different observables. 

We demonstrated the potential of our NEMol ansatz to describe the coupled nuclear and electron dynamics in molecular systems beyond diatomics. In \ce{NO2} we followed the dynamics in the excited state dominated by fast changing wavepacket interference effects. The ansatz is expandable to simulate the induced coherent electron dynamics in the excitation process itself as well as higher-dimensional molecular system as long as the underlying nuclear dynamics can be treated quantum mechanically. Two electron processes could be realized by using pair densities.   

% -----------------------------------------------------------------------------
\section*{Supplementary material}
See the supplementary material for the details of the wavepacket simulation setup, the underlying quantum chemical data of \ce{NO2} and additional figures and tables for the NEMol-dynamics. A section contains the results for the CEP-control obtained with the $y$-component and the $z$-component of the TDM. 
%Animations of the coupled electron density in the \mbox{1e-2o} picture for the free propagation and in the presents of a few-cycle IR laser pulse with a CEP of $0.0\pi$ and $1.4\pi$ are added in the supplementary material, too. 

\section*{Author contributions statement}
TS performed all calculations. TS and RDVR analyzed the results and contributed equally to the final version of the manuscript.

\section*{Data availability statement}
The data that support the findings of this study are available from the corresponding author upon reasonable request. The following article has been submitted to 'The Journal of Chemical Physics'. 

\section*{Conflicts of interest}
There are no conflicts to declare.

\section*{Acknowledgements}
The authors gratefully acknowledge the DFG Normalverfahren and the Munich Center of Advanced Photonics (MAP). 

% -----------------------------------------------------------------------------
% APPENDIX
% -----------------------------------------------------------------------------
\appendix*
\section{\label{sec:appendix}}
The following detailed formulation of the NEMol ansatz\cite{Geppert:2008,vondenHoff:2009,Znakovskaya:2009} is given here in the improved notation. The total molecular wavefunction $\Psi_{tot}(r,R,t)$ is setup as the sum over the electronic states with $\chi(R,t)$ the nuclear wavefunctions, $\varphi(r,t;R)$ the electronic wavefunctions, the nuclear and electronic coordinates $R$ and $r$ and the time $t$. 
\begin{equation}
\Psi_{tot}(r,R,t) = \sum_{i}  \chi_{i}(R,t) \cdot \varphi_{i}(r,t;R).
\end{equation}
Applying the Born-Oppenheimer approximation the uncoupled electronic wavefunctions $\varphi_{i}$ are hereby parametrically depending on the nuclear coordinates $R$ and define a multi-dimensional vector $\varphi_{tot}$. The total nuclear wavefunction $\chi_{tot}$ also represents a multi-dimensional vector, spanned by the coupled wavefunctions $\chi_{i}$. For details how the temporal evolution of the nuclear wavefunctions $\chi_{i}$ on coupled potential energy surfaces (PES) is determined see section~I of the SI. Multiplying $\Psi_{tot}(r,R,t)$ from the left with $\chi_{tot}$ and the subsequent integration over the nuclear coordinates results in an expression of the coupled total electronic wavefunction\cite{Geppert:2008,vondenHoff:2009,Znakovskaya:2009}.
\begin{align}
\begin{split}
\Phi_{tot}(r,t; \langle R \rangle(t))  & =  \int \chi_{tot}^{*}(R,t)  \cdot \Psi_{tot}(r,R,t) dR \\
 & =  \begin{pmatrix} \Phi_{1}(r,t; \langle R \rangle(t) ) \\ \Phi_{2}(r,t;\langle R \rangle(t)) \\ \vdots \\ 
 \Phi_{j}(r,t;\langle R \rangle(t) ) \end{pmatrix},
 \end{split}\\
 \text{with }  \langle R \rangle(t) & =   \sum_{i} \expval{R}{\chi_{i}(R,t)}_{R}. 
\end{align}
The coupled total electronic wavefunction is parametrically depending on the time-dependent expected value of the position $\langle R \rangle(t)$. In other words $\Phi_{tot}$ is evaluated at one single nuclear geometry which changes with time. The individual components $\Phi_{j}$ are defined by the following equation:
\begin{align}
\label{eq:coupledWF_ap}
\begin{split}
\Phi_{j}(r,t;\langle R \rangle(t)) & =  A_{jj}(t)\cdot  \varphi_{j}(r,t;\langle R \rangle(t)) \\
& +  \sum_{k \neq j } A_{jk}(t) \cdot  \varphi_{k}(r,t;\langle R \rangle(t)), 
\end{split}\\
\text{with } A_{jk}(t) & = \bra{\chi_{j}(R,t)}\ket{\chi_{k}(R,t)}_{R}.  
\end{align}
The first part depends on the population $A_{jj}$ of the respective state $j$, while all others summands include the nuclear overlap term $A_{jk}$ which specifies the degree of coherence induced between the two states $j$ and $k$. The population and coherence of the electronic states as well as the influence of all coupling terms are already determined by the nuclear quantum-dynamics simulation. If the coupling between the electronic states is weak, the nuclear wavefunctions propagate independently and the coherence term becomes zero. In this case, the coupled electronic wavefunctions $\Phi_{j}$ in equation~\ref{eq:coupledWF_ap} become equivalent to the uncoupled electronic wavefunction $\varphi_{j}$.
Standard quantum-chemical calculations at the $\langle R \rangle(t)$ structure yield the real-valued wavefunctions $\varphi_{j}(r;\langle R \rangle(t))$ of the relevant electronic states and their eigenenergies. The temporal evolution of $\varphi_{j}(r,t;\langle R \rangle(t))$  is determined by the deformation of the electronic structure induced by the nuclear motion (Born-Oppenheimer part) and an oscillation through phase space defined by a pure electronic phase.\cite{Geppert:2008,vondenHoff:2009,Znakovskaya:2009}
\begin{equation}
\varphi_{j}(r,t;\langle R \rangle(t)) = \varphi_{j}(r;\langle R \rangle(t)) \cdot e^{-i \xi_j(t)}
\end{equation}
The phase term $\xi_j(t)$ depends on the eigenenergies $E_{j}(\langle R \rangle(t))$ and has to be
calculated recursively. 
\begin{equation}
\xi_j(t)  =  E_{j}(\langle R \rangle(t)) \Delta t + \xi_{j}(t - \Delta t ).
\end{equation}
This recursive evaluation is necessary to retain the memory of the progressing electronic phase. Thereby the propagation velocity of the phase in the complex plane changes smoothly in time while the nuclear wavepacket propagates. Using the coupled total electronic wavefunction $\Phi_{tot}(r,t;\langle R \rangle(t))$ the associated electron density $\rho(r,t;\langle R \rangle(t))$ can be determined by multiplying $\Phi_{tot}(r,t;\langle R \rangle(t))$ from the left with $\varphi_{tot}$ and the subsequent integration over $N-1$ electronic coordinates (with $N$ being the total number of electrons).
\begin{widetext}
\begin{align}
\label{eq:coupledDens_ap}
\rho(r,t;\langle R \rangle(t))  =  \int \varphi_{tot}^{*} \cdot \Phi_{tot} dr_2 \dots dr_N
  & =  \sum_j  A_{jj}(t)\rho_{jj}(r;\langle R \rangle(t))  
  + \sum_{k \neq j } 2 Re \big\{  A_{jk}(t)\rho_{jk}(r;\langle R \rangle(t))e^{-i \xi_{jk}(t)} \big\}, \\
\text{with } \xi_{jk}(t) & =  \Delta E_{jk}(\langle R \rangle(t)) \Delta t + \xi_{jk}(t - \Delta t ).
\end{align}
\end{widetext}
% text erweitern
The first summation consists of the state specific electronic density $\rho_{jj}(r,t;\langle R \rangle(t))$ weighted with the corresponding time-dependent population $A_{jj}(t)$. The dynamics of these contributions to the coupled electron density is determined by the temporal evolution of the nuclear wavepacket i.e. its expected value of the position $\langle R \rangle(t)$. The second summation defines the coherent contribution to the coupled electron density and consists of the time-dependent overlap $A_{jk}(t)$, the one-electron transition density $\rho_{jk}(r,t;\langle R \rangle(t))$ and its pure electronic phase defined by the energy difference $\Delta E_{jk}$ between the involved electronic states.

\nocite{*}
% -----------------------------------------------------------------------------
% BIBLIOGRAPHY
% -----------------------------------------------------------------------------
\bibliography{main.bib}% Produces the bibliography via BibTeX.

\end{document}